\documentclass[a4paper,runningheads]{llncsreport}
\usepackage{orcidlink}
\usepackage[utf8]{inputenc}
\usepackage[T1]{fontenc}
\usepackage{llncsstuffthm}
\usepackage{cwdefs}
\usepackage{pgfkeys}
\usepackage{amssymb}
\usepackage{amsfonts}
\usepackage{amsmath}
\usepackage{bm}
\usepackage{multirow}
\usepackage{longtable}
\usepackage{booktabs}
\usepackage{bigdelim}
\usepackage{comment}
\usepackage{xspace}
\usepackage{booktabs}
\usepackage{graphicx}
\usepackage{upgreek}
\usepackage{enumerate}
\usepackage{rotating}
\usepackage{array}
\usepackage{xcolor}
\usepackage{colortbl}
\usepackage[caption=false]{subfig}
\usepackage{pdfpages}
\usepackage{wrapfig}
\usepackage{placeins} %
\usepackage{fancyvrb}
\usepackage{pgfplots}
\pgfplotsset{compat=1.18}
\usepackage{color}
\usepackage{hyperref}
\hypersetup{
hyperfootnotes=false,
breaklinks=true,
hidelinks,
linktoc=all,
colorlinks=false
}

\newcommand{\SWIPL}{\textit{SWI-Prolog}\xspace}

\newcommand{\TPTP}{\textit{TPTP}\xspace}

\newcommand{\Vampire}{\textit{Vampire}\xspace}
\newcommand{\SGCD}{\textit{SGCD}\xspace}
\newcommand{\EProver}{\textit{E}\xspace}
\newcommand{\EProverAS}{\textit{E-as}\xspace}
\newcommand{\EProverASC}{\textit{E-as-10}\xspace}

\newcommand{\ProverN}{\textit{Prover9}\xspace}
\newcommand{\leanCoP}{\textit{leanCoP}\xspace}

\newcommand{\CDTools}{\textit{CD Tools}\xspace}

\newcommand{\MM}{\name{Metamath}\xspace}

\newcommand{\oimp}{\Rightarrow}
\newcommand{\oimpn}{{\Rightarrow}}
\newcommand{\reduces}{\rightarrow}
\newcommand{\defpatt}{:=}
\renewcommand{\c}[1]{\mbox{\textsf{\textbf{#1}}}}
\renewcommand{\c}[1]{\bm{\mathsf{#1}}}
\newcommand{\ldot}{{\,.\,}}

\newcolumntype{A}{>{$}l<{$}}
\newcolumntype{B}{>{$}r<{$}}

\newcolumntype{L}[1]{>{\raggedright\let\newline\\\arraybackslash\hspace{0pt}}p{#1}}
\newcolumntype{C}[1]{>{\centering\let\newline\\\arraybackslash\hspace{0pt}}p{#1}}
\newcolumntype{R}[1]{>{\raggedleft\let\newline\\\arraybackslash\hspace{0pt}}p{#1}}
\newcolumntype{M}[1]{>{\raggedleft\let\newline\\\arraybackslash\hspace{0pt}$}p{#1}<$}

\renewcommand{\P}{\f{P}}
\newcommand{\D}{\f{D}}
\newcommand{\G}{\f{G}}

\newcommand{\Ax}{\mathit{Ax}}
\newcommand{\Patterns}{\mathit{Patterns}}
\newcommand{\Schemas}{\mathit{Schemas}}

\newcommand{\PSP}{\f{P}}
\newcommand{\TSIZE}{\f{T}}
\newcommand{\CSIZE}{\f{C}}

\newcommand{\HEIGHT}{\f{H}}
\newcommand{\PP}{\f{PP}}
\newcommand{\PK}{\f{PK}}

\renewcommand{\PSP}{\mathcal{P}}
\renewcommand{\TSIZE}{\mathcal{T}}
\renewcommand{\CSIZE}{\mathcal{C}}

\renewcommand{\HEIGHT}{\mathcal{H}}
\renewcommand{\PP}{\mathcal{PP}}
\renewcommand{\PK}{\mathcal{PK}}

\newcommand{\tsize}{\f{tsize}}
\newcommand{\csize}{\f{csize}}

\newcommand{\height}{\f{height}}
\newcommand{\saveval}{\f{sav}}
\newcommand{\refs}{\f{ref}}

\newcommand{\ST}{\f{Subterms}}

\newcommand{\hparen}{\hspace{0.5em}}

\newcommand{\h}[1]{\textit{#1}} %

\newcommand{\code}[1]{\texttt{#1}}
\newcommand{\fn}{\f{n}}

\newcommand{\mm}[1]{\texttt{#1}}

\newcommand{\pro}{\rightarrow}
\newcommand{\SETMM}{\textit{set.mm}\xspace}

\newcommand{\MGT}{\f{MGT}}

\newcommand{\TI}{T_{\mathit{Ideal}}}

\newcommand{\OEISNUM}[1]{\href{https://oeis.org/#1}{oeis:#1}}

\newcommand{\poi}[1]{\texttt{#1}}
\newcommand{\tptp}[1]{\texttt{#1}}

\renewcommand{\pl}[1]{\texttt{#1}}  %

\newcommand{\Abandoned}{\mathit{Abandoned}}
\newcommand{\hse}{\hphantom{ e}}

\newcommand{\NLem}{\xpar{NLem}}

\newcommand{\xpar}[1]{\textsc{#1}\xspace}

\newcommand{\xval}[1]{\texttt{#1}\xspace}

\newcommand{\POI}{\textit{POI}\xspace}
\newcommand{\POIBASE}{\textit{POI-base}\xspace}

\newcommand{\RID}{RunId}

\newcommand{\DLEM}{D_{Lem}}

\newcommand{\DIRECT}{\textit{d}\xspace}

\pgfplotscreateplotcyclelist{prec}{
    {red!90!black},
    {orange!60!black,dashed},
    {blue!90!black,dotted},
}

\newcommand{\Rbase}{\mathit{B}}
\newcommand{\RBP}{\mathit{BP}}
\newcommand{\RBT}{\mathit{BT}}
\newcommand{\RBPF}{\mathit{BPF}}
\newcommand{\RS}{\mathit{S}}
\newcommand{\RC}{\mathit{C}}

\newcommand{\RSBP}{\mathit{SBP}}
\newcommand{\RMBP}{\mathit{MBP}}
\newcommand{\RLBP}{\mathit{LBP}}
\newcommand{\RSBT}{\mathit{SBT}}
\newcommand{\RMBT}{\mathit{MBT}}
\newcommand{\RLBT}{\mathit{LBT}}

\newcommand{\LMS}{\mathit{LMS}}
\newcommand{\LMM}{\mathit{LMM}}
\newcommand{\LML}{\mathit{LML}}
\newcommand{\LMT}{\mathit{LMT}}
\newcommand{\LAL}{\mathit{A50}}
\newcommand{\LACP}{\mathit{A100P}}
\newcommand{\LACT}{\mathit{A100T}}

\newcommand{\STP}{\makebox[0.68em][c]{\textsf{P}}}
\newcommand{\STT}{\makebox[0.68em][c]{\textsf{T}}}
\newcommand{\STS}{\makebox[0.68em][c]{\textsf{S}}}
\newcommand{\STC}{\makebox[0.68em][c]{\textsf{C}}}
\newcommand{\STL}{\makebox[0.68em][c]{\textsf{L}}}
\newcommand{\STE}{\makebox[0.68em][c]{}}

\newcommand{\nocontentsline}[3]{}
\let\origcontentsline\addcontentsline
\newcommand\stoptoc{\let\addcontentsline\nocontentsline}
\newcommand\resumetoc{\let\addcontentsline\origcontentsline}

\newcommand{\appref}[1]{App.~\ref{#1}}
\newcommand{\apprefparen}[1]{(App.~\ref{#1})}

\newcommand{\tablesettings}
           {\fontsize{7.8pt}{8.2pt}\selectfont
             \renewcommand{\arraystretch}{0.9}}

           \newcommand{\eqdeftab}{:=}

\newcommand{\xsubsection}[1]{\subsection{#1}}

\begin{document}

\title{Generating Theorems by\\ Generating Proof Structures\\ (Extended
  Version)\\[-6pt]} \author{Christoph Wernhard}
\titlerunning{Generating Theorems by Generating Proof Structures}
\institute{University of Potsdam, Germany}

\maketitle

\begin{abstract}
  We address generating theorems from a given set of axioms, without proof
  goal, aiming at value from a mathematical point of view or as lemmas for
  automated proving. As benchmark, we convert a fragment of the Metamath
  database set.mm. Our techniques are centered on proof terms and condensed
  detachment. This ties in with automated first-order proving by proof
  structure enumeration, and links to Metamath and formulas-as-types. Our
  methods for generating theorems are based on partitioning the set of proof
  terms into inductively characterized levels. We study two ideas for
  improvement: Lemma synthesis by DAG compression of proof term sets, and
  incorporating combinators into proof terms. Our lemmas significantly improve
  solution rates of provers, e.g., of Vampire from 74\% to 94\%, and of
  leanCoP from 7\% to 44\%.
\end{abstract}

\stoptoc

\section{Introduction}
\label{sec-intro}

We address generating theorems, together with proofs, from a given set of
axioms, without proof goal, aiming at value from a mathematical point of view
or as lemmas for automated proving. This occurs as subtask in automated
proving and is on its own a classic objective of automated reasoning, e.g.,
with \name{Scott's challenge} \cite{wos:resonance:95}. Modern variations
include lemma generation \cite{rwzb:lemmas:2023}, premise selection, and, more
generally, learning and AI methods for automated reasoning
\cite{guidedatp:survey:2025}.
Given an ideal solution set $\TI$, the task has two aspects: The \name{recall
  aspect}, finding a set of theorems that contains as much members of $\TI$ as
possible, and the \emph{precision aspect}, finding a set of theorems that has
as few members as possible that are not in $\TI$. We consider here mainly the
recall aspect.

Our techniques are based on proof terms and condensed detachment. On the one
hand this links to \MM \cite{megill:1995} and formulas-as-types; on the other
hand it ties in with proof search through enumeration of proof structures,
interwoven with formula unification, as in the connection method
\cite{bibel:atp:1987,bibel:otten:2020} and clausal tableaux \cite{letz:habil}.
Redundancies are avoided by having each structure (proof) only once in the
enumeration, and by eliminating sets of structures early on through failure of
unification for a substructure.
Commonly, e.g., with \leanCoP \cite{leancop}, this is employed ``top-down'',
utilizing unifiability through a ground-instantiated goal theorem. Compared to
saturation-based provers, a serious drawback is that proven subgoals are not
re-used in \emph{fresh copies}. However, proof structures can also be
enumerated ``bottom-up'' without a given goal
\cite{schumann:delta:1994,rwzb:lemmas:2023,cw:sgcd:2024}, and as DAGs with
copying formulas for each incoming edge \cite{eder:cs:1989,cw:ccs:2022}. We
utilize this for generating theorems by generating proof terms, based on
\name{inductive level characterizations}.

As a benchmark $\TI$, we convert a fragment of 1,374 theorems from the \MM
database \SETMM \cite{metamath:book,metamath:setmm}. We identify 554 of
these as easy to obtain by generating proof terms. With two novel techniques,
we generate proofs for 180 further ones: (1)~Lemma synthesis based on the
hypothesis that a formula is a valuable lemma if it is proven by a proof term
with significant compressing effect on a set of previously generated proof
terms. (2)~Incorporating combinators\footnote{In the sense of combinatory
logic \cite{schoenfinkel:24:bausteine,curry:1958}. For a practice-oriented
introduction see \cite{peytonjones:87}.} into proof terms for their
compressing effect.
Several of our generated theorems are hard to prove for first-order provers.
Supplied to provers, our lemmas significantly improve solution rates, e.g.,
for \Vampire \cite{vampire} from 74\% to 94\%, and for \leanCoP \cite{leancop}
from 7\% to 44\%.

After preliminaries (Sect.~\ref{sec-bg}), we describe our benchmark
(Sect.~\ref{sec-ideal}) and develop our techniques for generating
theorems/proof terms (Sect.~\ref{sec-generating-proof-structures}). Then we
identify the easy-to-generate theorems (Sect.~\ref{sec-basesolved}), and study
our improvements by lemma synthesis (Sect.~\ref{sec-dag-synthesis}) and
incorporation of combinators (Sect.~\ref{sec-combinatory}).
Section~\ref{sec-conclusion} concludes the main part of this report, which
extends the conference publication \cite{cw:generating:2026} by appendices
with supplementary material.
The presented techniques are implemented in \SWIPL
\cite{swiprolog} with \CDTools \cite{cw:cdtools:website}. Artifacts are
accessible from a dedicated Web page \cite{cw:thgen:web}.

\enlargethispage{0.3cm}
\vspace{-0.1cm}
\section{Technical Framework: Condensed Detachment}
\label{sec-bg}

\begin{table}[t]
  \centering
  \caption{The axiom system of \SETMM for classical propositional logic.}
  \label{tab-propaxioms}
  \begin{tabular}{cl@{\hspace{1em}}l@{\hspace{1em}}l}
    \textit{Identifier} & \textit{Axiom} & \textit{Names} \cite{ulrich:legacy:2001}\\\midrule
    1 & $x\oimp (y\oimp x)$ & $\c{K}$, \textit{Simp}\\
    2 & $(x\oimp (y\oimp z))\oimp ((x\oimp y)\oimp (x\oimp z))$ & $\c{S}$, \textit{Frege}\\
    3 & $(\fn(x)\oimp \fn(y))\oimp (y\oimp x)$ & \textit{Transp}
  \end{tabular}
  \vspace{-15pt}
\end{table}

\name{Condensed detachment (CD)}, in essence modus ponens with unification,
was introduced by Carew A. Meredith in the 1950s \cite{prior:logicians:1956}.
With proof terms, it can be viewed from the perspective of formulas-as-types
\cite{hindley:meredith:cd:1990,hindley:book:1997,rezus:2020:witness}. In
automated proving, it led to numerous advances up to the early 2000s
\cite{ulrich:legacy:2001}, and recently its relevance for approaches based on
proof structures was observed \cite{cwwb:lukas:2021,cwwb:article:2024}. It is
the basis of \MM \cite{megill:1995} with its database \SETMM
\cite{metamath:setmm}.
It is often applied to axiomatizations of propositional logics. From the
first-order point of view, this involves a single unary predicate $\P$
(``\name{provable}''), whereas the propositional \defname{object-level
  formulas} are represented as terms. We consider those defined by the
grammar
\[f,g\; :=\; x \mid f \oimp g \mid \fn(f),\]
where $x$ ranges over a set of \defname{variables} $x,y,z,u,v,w$, also with
index subscript. From the first-order point of view, $\oimp$ is a binary
function symbol representing object-level implication, and $\fn$ is a unary
function symbol representing object-level negation.
Table~\ref{tab-propaxioms} shows the axioms of classical propositional logic
in \SETMM, written as object-level formulas. From the first-order perspective,
e.g., axiom~1 is $\forall x \forall y\, \P(x\oimp (y\oimp x))$.
CD \defname{proof terms} are defined by the grammar
\[d,e\; :=\; a \mid \D(d,e),\]
where $a$ ranges over a finite set of constants, called \defname{axiom identifiers}.
With each axiom identifier $a$ an object-level formula $f_a$ is associated.
The \defname{most general theorem (MGT)} of a proof term is a positive
unit clause defined as follows.
\begin{itemize}
\item $\MGT(a) \eqdef \P(f_a)$, if $a$ is an axiom identifier;
\item $\MGT(\D(d_1,d_2))$ is the unit clause resulting from the clause
  \begin{equation}
    \tag{\textit{Det}}
    \label{eq-det}
    \P(y) \revimp (\P(x \oimp y) \land \P(x)),
  \end{equation}
  by two resolution steps: upon the body atom $\P(x \oimp y)$ with
  $\MGT(d_1)$, and upon the obtained instance of the second body atom $\P(x)$
  with $\MGT(d_2)$.
\end{itemize}
As usual, we assume that variables of resolved clauses are renamed such that
they do not share variables. If $\P(f)$ is the MGT of some proof term, we
sometimes also call just $f$ the MGT. A proof term may have no MGT (w.r.t. a
mapping of axiom identifiers to object-level formulas) since literals to be
resolved upon may not unify. The MGT is then said to be \name{undefined}. If
defined, the MGT of a proof term is unique up to renaming of variables. The
MGT is the \name{most general} formula proven by a proof term since the
formulas it proves are the MGT itself and all its substitution instances. For
more precise discussions see
\cite{cwwb:article:2024,cw:zz:2025,hindley:meredith:cd:1990,hindley:book:1997}.\footnote{In
the formulas-as-types view $\D$ is \name{application} and the MGT is the
\name{principal type}.}

\section{\POI: A Benchmark Theorem Set from \SETMM}
\label{sec-ideal}

As benchmark set for generating theorems we extract those theorems from the
core part of \SETMM\footnote{\url{https://github.com/metamath/set.mm}, from
July 5, 2025 (commit 011fba3).} that are proven from its axioms for classical
propositional logic (Table~\ref{tab-propaxioms}). We call this set of theorems
\defname{\POI} (for \name{propositional-only implication form}). We start from
the first two thirds of \SETMM with 28,673 curated theorems about various
topics, dropping the remainder of deprecated or unstable material, and select
those theorems whose proof depends (aside of $\D$) only on axioms
$1,2,3$\footnote{Axioms 1,2,3 are \mm{ax-1}, \mm{ax-2}, \mm{ax-3} in \SETMM.
$\D$ is \mm{ax-mp} with switched arguments.} and special axioms introducing
object-level symbols.
To facilitate use with external systems, we canonicalize the theorem formulas
as follows.

\begin{enumerate}
\item \textit{Conversion to implication form.}
  \label{conv-step-imp-form}
  From a first-order point of view, \MM theorems are definite clauses $\P(g)
  \revimp (\P(f_1) \land \ldots \land \P(f_n))$
  \cite{cw:zz:2025,metamath:atp:2023}, which we convert to unit clauses
  $\P(f_1 \oimp (f_2 \oimp \ldots (f_n \oimp g)))$. Proofs for both forms can
  be linearly translated into one another \apprefparen{app-implication-form}.
  \item \textit{Eliminating defined symbols.} In \SETMM object-level symbols
    other than $\oimp$ and $\fn$ are introduced with ``definitional'' axioms,
    recognizable by the name prefix \mm{df-}. We unfold these definitions. In
    proofs of \SETMM, the \mm{df-} axioms can be replaced with proofs of
    associated theorems.
  \item \textit{Removing duplicates.} If, after these conversions, formulas
    for different theorems from \SETMM are identical, we only record the name
    of the first encountered, comparing first with theorems that are unit
    clauses already in \SETMM. We keep theorems that are strictly subsumed by
    others, since a subsumed theorem may be provable with a shorter proof.
\end{enumerate}

As benchmark set \POI we finally obtain a set of 1,374 theorems. Their
object-level formulas are valid formulas of classical propositional logic. For
58\% of them the source was a definite clause with nonempty body.

\begin{wraptable}[9]{r}{4cm}
  \vspace{-1.2cm}  
\caption{Prover results.}
\label{tab-prover-results}\small
\vspace{4pt}
\begin{tabular}{lrrr}
\h{Prover} & \h{3,600 s} & \h{600 s} & \h{60 s}\\\midrule
\Vampire & 1,028 & 952 & 571\\
\EProver & 911 & 820 & 664\\
\EProverAS & 868 & 826 & 669\\
\ProverN & 314 & 303 & 258\\
\leanCoP & 92 & 84 & 76\\\midrule
\textit{Total} & 1,038 & 988 & 740
\end{tabular}
\end{wraptable}
For each POI theorem we provide a \TPTP \cite{tptp} \code{FOF} problem. Its
axioms are the universal closures of (\ref{eq-det}) and, as first-order atoms
with predicate $\P$, the axioms from Table~\ref{tab-propaxioms}. The
conjecture is the theorem formula, as universally closed atom with predicate
$\P$.
To estimate difficulty for first-order provers, we tried \Vampire
\cite{vampire}, \EProver \cite{eprover}, \ProverN \cite{prover9} (which stands
out because its proofs can be converted to our proof terms), and \leanCoP
\cite{leancop}. Table~\ref{tab-prover-results} shows how many of the 1,374
problems were solved in 3,600, 600, and 60
seconds\footnote{\label{foot-cluster}Versions and settings: \Vampire 5.0.0,
\code{--mode casc}; \EProver 3.2.5, \code{--auto}; \EProverAS is \EProver
3.2.5, \code{--auto-schedule}; \ProverN \code{auto}; \leanCoP 2.1b. Time
limit: 3,600~s. Platform: Intel Xeon 9242 cluster. Best results of two runs,
with RAM limits 3.77~GB and 15~GB.} (details: \appref{app-bigtable}).
We use the results of \Vampire, \EProver and \EProverAS to assign a
``difficulty rating'' to each \POI theorem, between 0 (most easy) and 1 (most
hard): For \POI theorems $p$ we define
$\f{rating}(p) \eqdef 1 - \sum_{s \in \{\text{\Vampire},
  \text{\EProver}, \text{\EProverAS}\}, t \in \{60,600,3600\}}
\f{solved}(p,s,t)/9$,
where $\f{solved}(p,s,t)$ is $1$ if the proving
problem for $p$ was solved by prover $s$ in our runs in at most $t$
seconds, and $0$ otherwise.

\section{An Approach to Generating Proof Structures}
\label{sec-inductive}

\label{sec-generating-proof-structures}

We develop a systematic approach to generate theorems by generating, or
enumerating, proof structures.

\xsubsection{Inductive Level Characterizations}

We assume a fixed finite set $\Ax$ of axiom identifiers, e.g., $\{1,2,3\}$
from Table~\ref{tab-propaxioms}.
For a proof term $d$ let $\ST(d)$ denote the (not necessarily proper) set of
all subterms of $d$ that are compound, i.e., not in $\Ax$. Define
\defname{tree size}, \defname{compacted size}, and \defname{height} of a proof
term as follows.
\begin{itemize}
\item $\tsize(a) \eqdef 0$ if $a \in \Ax$; $\tsize(\D(d,e)) \eqdef 1 + \tsize(d) + \tsize(e)$.
\item $\csize(d) \eqdef |\ST(d)|$.
\item $\height(a) \eqdef 0$ if $a \in \Ax$; $\height(\D(d,e)) \eqdef 1 +
  \f{max}(\height(d),\height(e))$.
\end{itemize}
The \name{compacted size} is also called \defname{DAG-size} since it equals
the number of inner nodes of the unique minimal DAG representing the proof
term.
\pagebreak

A \defname{level characterization} $L$ is a specification of a partition of
the set of all proof terms into a series $L_0, L_1, L_2, \ldots$ of disjoint
subsets, called \emph{levels}.
Each measure defined above can be taken as basis for a level characterization:
$\TSIZE_i \eqdef \{d \mid \tsize(d) = i\}$, $\CSIZE_i \eqdef \{d \mid
\csize(d) = i\}$, $\HEIGHT_i \eqdef \{d \mid \height(d) = i\}$.
Using level characterizations to organize proving or generating theorems is
made possible by \emph{inductive level characterizations}, which for $\TSIZE$
and $\HEIGHT$ can be given as follows.
\begin{defn}[Inductive Level Characterizations: Tree Size and Height]
  \label{def-indu-th}
  \noindent
  $\begin{array}{lcl}
  \TSIZE_0 & \eqdef & \Ax;\\
  \TSIZE_{n+1}\!\! & \eqdef & \{\D(d,e) \mid \exists i{\leq}n\; .\; d \in
  \TSIZE_i,\; e \in \TSIZE_{n-i}\}.\\[0.5ex]
  \HEIGHT_0 & \eqdef & \Ax;\\
  \HEIGHT_{n+1}\!\! & \eqdef & \{\D(d,e) \mid d \in \HEIGHT_n,\; \exists i {\leq} n\;.\; e \in \HEIGHT_i\}
    \cup \{\D(d,e) \mid e \in \HEIGHT_n,\; \exists i{<}n\;.\; d \in \HEIGHT_i\}.
  \end{array}$
\end{defn}
An inductive level characterization can be read in two ways as high-level
specification of an algorithm to generate proof terms level-by-level, e.g.,
for~$T$: (1)~Top-down with iterative deepening: For increasing $n$, proof
terms $\D(d,e)$ in $\TSIZE_{n+1}$ are generated by generating proof terms~$d$
in levels $T_i$ for $i \leq n$, and, for each $d$, generating proofs~$e$ in
level $T_{n-i}$. (2)~Bottom-up: For increasing $n$, proofs in $\TSIZE_{n+1}$
are generated and cached. When generating proof terms $\D(d,e)$ in
$\TSIZE_{n+1}$, suitable $d$ and $e$ are retrieved from cached levels $T_i$
and $T_{n-i}$ with $i \leq n$.

For proving or generating theorems, generation of levels is interwoven with
MGT computation. Proof terms without defined MGT are excluded from computed
levels. A proof term only has a defined MGT if all its subterms have one.
Thus, if strict subterms of proof terms are in lower levels, the reducing
effect of requiring defined MGTs transfers from lower to higher levels.

Bottom-up processing based on inductive level characterizations offers two key
features of importance for proving and generating theorems. (1)~\emph{Caching}
to avoid recomputations: When generating a level, proof terms that are
subterms and in a lower level are retrieved from the cached lower level and
not computed anew. Through copying MGTs of re-used subproofs, proof terms can
be represented by DAGs instead of trees. (2)~A certain \emph{simple way to
incorporate restrictions}: Computed levels can be restricted to proof terms
that meet heuristic criteria. If strict subterms of proof terms are in lower
levels, the reduction is passed on from lower to higher levels: a lower level
with fewer proof terms makes fewer ones available as strict subterms of proof
terms in higher levels. Useful heuristic restrictions can refer to MGTs, e.g.,
with a formula size limit. They may depend on the order in which proof terms
are generated, e.g., if proof terms whose MGT is subsumed by that of a
preceding one are to be excluded. While these features are common for
saturation-based provers, they are foreign to conventional
structure-generating provers, which are proceeding top-down.

\xsubsection{Considering Compacted Size}
\label{subsec-compacted}
Level characterization by \emph{compacted size} is appealing for two
reasons. First, the compacted size of a proof term is often significantly
smaller than its tree size. Let
$d_1 \eqdef \D(\D(1, 1), \D(\D(1, \D(1, 1)), \D(1, \D(1, 1))))$,
which has tree size~$7$ and compacted size~$4$. DAG grammars\footnote{I.e.,
tree grammars \cite{lohrey:survey:2015} that are non-cyclic, have a single
production for each nonterminal, and have only nonterminals without
parameters.} provide a term notation that respects compacted size:
\begin{equation}
  \label{eq-indu-d1-dag}
  \{4 \pro \D(1, 1),\; 5 \pro \D(1, 4),\; \mathit{Start} \pro
  \D(4, \D(5, 5))\}.
\end{equation}
As nonterminals, we use $\mathit{Start}$ and numbers larger than the axiom
identifiers $1,2,3$. In accord with compacted size, we find 4 occurrences of
$\D$ in right-hand sides (RHSs) of the productions. Proofs in \SETMM often
have astronomical tree size but modest compacted size, e.g., 2.53$\times
10^{22}$ versus $3,647$ for theorem \mm{peano3}.

The second motive for level characterization by compacted size is
incorporating \name{combinators} like axioms into proof term enumeration
\cite{cw:ccs:2022}. Combinators restructure proof terms such that new
possibilities of factoring shared subterms emerge, reducing the compacted
size. At the same time, they can be handled in determining MGTs like ordinary
axioms, with their principal type as axiom formula.
As example consider the proof term
\begin{equation}
  \label{eq-indu-d2}
  d_2 \eqdef \D(\D(\D(\c{B}, 1), 1),
  \D(\D(\D(\c{B}, 1), 1), \D(\D(\D(\c{B}, 1), 1), 1))).
\end{equation}
We can write $d_2$ as DAG grammar, annotated with MGTs of the expanded
subterms, headed by rows for $1$ and $\c{B}$:
\begin{equation}
  \label{eq-indu-d2-dag}
\begin{array}{rclcl}
    1 & & & : & x\oimpn (y\oimpn x)\\
\c{B} & & & : & (x\oimpn y)\oimpn ((z\oimpn x)\oimpn (z\oimpn y))\\
4 & \pro & \D(\D(\c{B}, 1), 1) & : & x\oimpn (y\oimpn (z\oimpn x))\\
\mathit{Start} & \pro & \D(4, \D(4, \D(4, 1))) & : &
x\oimpn (y\oimpn (z\oimpn (u\oimpn (v\oimpn (w\oimpn (x_1\oimpn (y_1\oimpn x_1))))))).
\end{array}\hspace{-1cm}
\end{equation}
Counting $\D$'s in RHSs, we see that the compacted size of $d_2$ is 5. We can
eliminate the $\c{B}$ combinator from $d_2$ by rewriting with its reduction
rule, i.e., $\D(\D(\D(\c{B}, V_1), V_2), V_3) \reduces \D(V_1,\D(V_2,V_3))$.
The result is the proof term
\begin{equation}
  \label{eq-indu-d2-expanded}
  d_3 \eqdef \D(1, \D(1, \D(1, \D(1, \D(1, \D(1, 1)))))),
\end{equation}
which has the same MGT as $d_2$ but the larger compacted size 6. Incorporating
combinators can effect exponential reductions of compacted size
\cite{cw:ccs:2022}.
Simulations of resolution by methods based on proof structures
\cite{eder:cs:1989,eder:relative:1992} can be understood as incorporating
combinators and enumeration by compacted size \cite{cw:ccs:2022}.

Generation of proof terms in levels for increasing compacted size can be
performed by an adaptation \cite{cw:ccs:2022} of the \name{value-number
  method} for DAG enumeration \cite{aho:compilers:86,genitrini:2020}.
Unfortunately, this does not provide an inductive characterization like those
of Def.~\ref{def-indu-th}, which give way to caching and incorporating
restrictions.
However, if we give up completeness, i.e., accept that there are proof terms
that are not in any $L_i$, the \defname{PSP} (for \name{Proof-SubProof})
\defname{level} \cite{cwwb:lukas:2021,cwwb:article:2024} is a suitable
inductive level characterization by compacted size:
\begin{defn}[Inductive Characterization of the PSP Level]
  \label{def-indu-p}
  \[\begin{array}{lcl}
  \PSP_0 & \eqdef & \Ax;\\
  \PSP_{n+1} & \eqdef &  \hphantom{\cup\;} \{\D(d,e) \mid d \in \PSP_n,\; e \in \ST(d) \cup \Ax \}\\
  && \cup\; \{\D(d,e) \mid e \in \PSP_n,\; d \in \ST(e) \cup \Ax \}.
  \end{array}\]  
\end{defn}
A proof term in PSP level $i$ has compacted size $i$, in other words: $\PSP_i
\subseteq \CSIZE_i$. For $i \leq 3$ it holds that $\PSP_i = \CSIZE_i$.
Generating proofs by PSP level was inspired by proofs from logicians
\cite{cwwb:lukas:2021,cwwb:article:2024}. So far it has been applied
\cite{rwzb:lemmas:2023,cwwb:lukas:2021,cw:sgcd:2024} to single-axiom CD
problems such as \tptp{LCL038-1} \cite{pfenning:single:1988} and the hard
problem \tptp{LCL073-1} \cite{wos:meredith}.

\xsubsection{Applying the \SGCD System}

The \SGCD (\name{Structure-Generating theorem proving for Condensed
  Detachment}) prover \cite{rwzb:lemmas:2023,cw:sgcd:2024}, a part of
\CDTools, implements methods based on inductive level characterizations. It is
highly configurable, supporting interaction of top-down and bottom-up
processing. Here we focus on bottom-up generation of theorems from given
axioms.
\SGCD can be configured for level characterizations $L \in \{\TSIZE, \HEIGHT,
\PSP\}$ as given in Defs.~\ref{def-indu-th} and~\ref{def-indu-p}.
It then computes for increasing $i \in \{0,1,2,\ldots\}$ a \defname{cached
  solution set} $M_i \subseteq \bigcup_{0}^{i} L_i$. $M_0 = \Ax$. For $i \geq
1$, first a set $M'_i \subseteq L_i$ is generated on the basis of $M_{i-1}$
such that whenever $d$ is a subterm of a member of $M'_i$ in a level $L_j$
with $j < i$, then $d \in M_{i-1}$. Generation of $M'_i$ can be constrained by
configurable restrictions \xpar{Rest}, discussed below. $M_i$ is formed from
$M'_i \cup M_{i-1}$: the set is ordered according to configured criteria
\xpar{Ord}, and the \xpar{Trim} (a natural number) top members are extracted
and postprocessed as configured with \xpar{Post}, to yield~$M_i$. \SGCD
maintains a further set of proof terms, $\Abandoned$, that records for all
considered~$i$ proof terms $(M'_i \cup M_{i-1}) \setminus M_i$.

\SGCD terminates with \xval{Exhausted} if no new proof terms are obtained for
$M'_i$, i.e., if $M'_i = M_{i-1}$. It terminates with \xval{Stopped}, if $M_i$
for a configured maximum~$i$, \xpar{MaxLevel}, has been computed. The result
of an \SGCD run consists of two proof term sets, the last $M_i$ and
$\Abandoned$. The cardinality of $M_i$ is at most \xpar{Trim}. For theorem
generation both sets are relevant.
The following vector summarizes key parameters of \SGCD for bottom-up
theorem generation:

\smallskip
{\centering $\la L, \xpar{Rest}, \xpar{Ord}, \xpar{Trim}, \xpar{Post},
  \xpar{MaxLevel}\ra.$\par}
\smallskip

\noindent For \xpar{Rest}, configurable settings include: \xval{dup}
(\xval{subs}): exclude proof terms whose MGT is identical to (subsumed by)
that of a proof term computed before for an $M'_j$ with $j \leq i$;
\xval{lt\_ft($n$)} (\xval{lt\_fh($n$)}): exclude proof terms whose MGT has
tree size (height) larger than $n$;\footnote{Size measures analogous to those
for proof terms, applied to object-level formulas.} \xval{gen\_max($n$)}:
Compute at most $n$ members of $M'_i$.
For \xpar{Ord}, configurable ordering criteria include lexical comparison by
MGT size values, smallest preferred, specifically \xval{f\_th}: by tree size
and then height, and \xval{f\_ht}: by height and then tree size.
An option for \xpar{Post} is \xval{subs}, subsumption reduction: remove
members with the same MGT except one, and remove members whose MGT is strictly
subsumed by that of another member.

\begin{table}[t]
  \centering
  \caption{Legend for tables summarizing \SGCD runs.}
  \label{tab-run-legend}
  \tablesettings
  
  \begin{tabular}{l@{\hspace{4pt}}L{33em}}
    \h{Column} & \h{Explanation}\\\midrule
    $\h{\RID}$ & Identifier of the \SGCD run.\\
    $L$ &  Level characterization.\\
    \h{\POI} & Number of ``solved'' \POI theorems, i.e., \POI theorems for which a
    proof is
    in the result set $M_i \cup \Abandoned$.\\
    \h{1} &  Number of ``solved'' \POI theorems with rating 1.\\
    \h{NB} &  Number of ``solved'' \POI theorems not in \POIBASE (defined below).\\    
    \xval{lt\_ft} & Configured value $n$ of the \xval{lt\_ft($n$)} restriction in \xpar{Res}.\\
    \h{GMax} & Configured value of the \xval{gen\_max} restriction.\\
    \xpar{Trim} & Configured value of the \xpar{Trim} parameter.\\
    \xpar{Ord} & Configured value of the \xpar{Ord} parameter.\\ \h{Lev} &
    Level $i$ of the final $M_i$. This was either configured as
    \xpar{MaxLevel} or \SGCD terminated with \xval{Exhausted},
    indicated by suffix ``e''.\\
    \h{Time} &  Wall clock time of the run in seconds.\\
    \h{Mem} &  Peak value of heap use after computing an $M'_i$ in GB.\\
    |\h{Gen}| &  Total number of generated proof terms, i.e., final value of $|M_i \cup
    \Abandoned|$.\\
  \end{tabular}

  \medskip
  
  \caption{Two runs of \SGCD that contribute to \POIBASE. \xpar{Rest} is set
    to \xval{dup} and \xpar{Post} to \xval{subs}. The last row refers to the
    union of the results of both runs.}
  \label{tab-base-pt}
  \centering
  \tablesettings
  \setlength{\tabcolsep}{4pt}
\begin{tabular}{llrrrrrrrr}
   \h{\RID} & $L$  & \h{\POI} & \h{1} & \xpar{Trim} & \xpar{Ord} & \h{Lev} & \h{Time} & \h{Mem} & $|\h{Gen}|$\\\midrule
   $\RBP$ & $\PSP$   & 494 & 4          & 200k & \xval{f\_th} & 64 e & 17,124 & 53 & 13,108k\\
   $\RBT$ & $\TSIZE$ & 462 & 1          & 20k & \xval{f\_ht} & 592 e & 19,226 & 123 & 39,114k\\\midrule
   \multicolumn{2}{l}{$\RBP,\RBT$}       & 544 & 4\\
\end{tabular}

\medskip

  \caption{Five runs of \SGCD that contribute to \POIBASE. \xpar{Res} is
    parameterized with \xval{lt\_ft($n$)} for $n$ between 8 and 12, and with
    $\xval{subs}$. \xpar{Trim} is unlimited. The last three rows refer to
    unions of runs.}
`  \label{tab-base-inc}
  \centering
  \tablesettings
  \setlength{\tabcolsep}{4pt}
\begin{tabular}{llrrrrrrr}
   \h{\RID} & $L$  & \h{\POI} & \h{1} & \xval{lt\_ft} & \h{Lev} & \h{Time} & \h{Mem} & \h{Gen}\\\midrule
   $\RBPF_8$ & $\PSP$ & 323 & 0 & 8 & 63 e & 2,494 & 3 & 64,937\\
   $\RBPF_9$ & $\PSP$ & 377   & 0 & 9 & 25\hse & 7,500 & 2 & 27,3766\\
   $\RBPF_{10}$ & $\PSP$ & 232 & 0 & 10 & 17\hse & 5,555 & 2 & 325,454\\
   $\RBPF_{11}$ & $\PSP$ & 228 & 0 & 11 & 15\hse & 4,638 & 2 & 324,717\\
   $\RBPF_{12}$ & $\PSP$ & 201 & 0 & 12 & 14\hse & 4,602 & 2 & 352,928\\\midrule
   \multicolumn{2}{l}{$\RBPF_{8-12}$} & 426 & 0\\
   \multicolumn{2}{l}{$\RBPF_{8-12},\RBP$} & 507 & 4\\
   \multicolumn{2}{l}{$\Rbase \eqdeftab \RBPF_{8-12},\RBP,\RBT$} & 554 & 4\\      
\end{tabular}
\vspace{-20pt}
\end{table}

\section{\textit{POI-Base}: An Easy-to-Generate Subset of \POI}
\label{sec-basesolved}

We define \POIBASE as the set of those theorems from \POI for which proofs
appear in outputs of a few specific runs of \SGCD representing straightforward
proof term enumerations, constrained by straightforward heuristic
restrictions, limited to a few hours on commodity hardware.\footnote{The
platform for all \SGCD runs described in the paper was \SWIPL~10.0.0 on a
Linux server with four Intel Xeon E5-4640 processors and 256 GB RAM.}
Table~\ref{tab-base-pt} summarizes a run with level characterization $\PSP$
and a run with $\TSIZE$. (For the legend of all tables summarizing \SGCD runs
see Table~\ref{tab-run-legend}.) Values for \xpar{Trim} were set to prevent a
quick growth of the time spent for the levels.

Taken together, the two runs identify 544 \POIBASE problems.
Table~\ref{tab-base-inc} summarizes a series of five further runs with level
characterization $\PSP$, restricted by limiting the tree size of the MGT.
We finally define \defname{\POIBASE} as the set of the 554 \POI theorems for
which proofs are obtained in at least one of the \SGCD runs $\RBP,\RBT$, and
$\RBPF_{8-12}$. We declare $\Rbase$ as identifier of a ``virtual run'' of
\SGCD that yields proofs for these 554 theorems.

Concerning the prover results from Sect.~\ref{sec-ideal} we observe the
following: Four \POIBASE theorems have rating 1, i.e., the corresponding
proving problems are not easy for either \Vampire or \EProver. For all 92
problems proven by \leanCoP and for all 314 proven by \ProverN the theorem
belongs to \POIBASE. For those proven by \leanCoP, proofs were generated by
just $\RBT$, by just $\RBP$, and by just $\RBP_{8-12}$.
For observations concerning proof term sizes see \appref{app-proofsize}.

\section{Lemma Synthesis via Proof Term Compression}
\label{sec-dag-synthesis}

We now pursue the hypothesis that a formula is a valuable lemma if it is
proven by a proof term with a significant compressing effect on some large set
$D_0$ of proof terms with defined MGT.

\xsubsection{Computing the Lemmas}
As proof term compression we consider DAG compression, a special case of
grammar-based tree compression \cite{lohrey:treerepair:2013,cw:zz:2025} with a
unique best compression, the minimal DAG, which can be efficiently
computed.\footnote{\SWIPL supports this with the library predicate
\code{term\_factorized}/3.} The size of a DAG grammar $G = \{a_i \pro d_i \mid
i \in \{1,\ldots,n\}\}$ is the sum of the tree sizes of its RHSs: $|G| \eqdef
\sum_{i = 1}^{n} \tsize(d_i)$.\footnote{For general grammar-based tree
compression the number of \emph{edges} is more appropriate as basis
\cite{lohrey:treerepair:2013,cw:zz:2025}. For our full binary trees this is
twice the number of inner nodes.} The compressing effect of a production $a_k
\pro d_k$ in $G$ is indicated by its \defname{save-value}
\cite{lohrey:treerepair:2013}, the difference of the size of $G$ after
unfolding the production (i.e., removing it and replacing $a_k$ by $d_k$ in
all other RHSs) and the original size of $G$. It can be calculated as
$\saveval_G(a_k \pro d_k) \eqdef (\refs_G(a_k) - 1) * \tsize(d_k)$, where
$\refs_G(a_k)$ is the number of occurrences of $a_k$ in the RHSs of $G$.

Our workflow for lemma synthesis is as follows. A base set $D_0$ of proof
terms is generated with \SGCD and compressed into its minimal DAG, represented
by a DAG grammar $G$. The productions of $G$ are ordered according to lexical
comparison of (1.) save-value (largest preferred), (2.) tree size of the MGT
(smallest preferred), and (3.) height of the MGT (smallest preferred). The
considered MGT for a production is the MGT of the full expansion of its RHS.
An implementation can utilize the grammar representation to avoid actual
expansion when computing these MGTs. Productions with save-value 0, i.e.,
``top-level'' productions whose LHS does not occur in a RHS, are then deleted.
The RHSs of the remaining productions are now fully expanded (efficiently
through structure sharing) to get a sequence $\DLEM'$ of proof terms, ordered
by expected usefulness of their MGTs as lemmas. The final sequence of proof
terms $\DLEM$ is obtained from $\DLEM'$ by subsumption reduction, i.e.,
removing members with the same MGT except one, and removing members whose MGT
is strictly subsumed by that of another member.
The MGTs of the prefix of $\DLEM$ with a configured length \NLem are then
given to applications as additional axioms in lemma role.

\begin{table}[t]
  \caption{\SGCD runs used to compose base sets $D_0$ in our experiments.}
  \label{tab-addaxioms-bases}
  \centering
  \tablesettings  
  \setlength{\tabcolsep}{2.5pt}
\begin{tabular}{lcrrrrrrrr}
   \h{\RID} & $L$ & \h{NB} & \h{\POI} & \h{1} &  \xpar{Trim} & \xpar{Ord} & \h{Lev} & \h{Time} & $|\h{Gen}|$\\\midrule   
$\RSBP$ & $\PSP$ & 0 & 232 & 0 & 5k & \xval{f\_th} & 55 e & 175 & 161k\\
$\RSBT$ & $\TSIZE$ & 0 & 277 & 0 & 1k & \xval{f\_ht} & 161 e & 138 &  256k\\
$\RMBP$ & $\PSP$ & 1 & 366 & 1 & 20k & \xval{f\_th}  & 67 e & 828 & 973k\\
$\RMBT$ & $\TSIZE$ & 0 & 392 & 1 & 3k & \xval{f\_ht} & 219 e & 714 & 1,926k\\
$\RLBP$ & $\PSP$ & 4 & 419 & 2 & 50k & \xval{f\_th} & 103 e & 4,565 & 2,574k\\
$\RLBT$ & $\TSIZE$ & 0 & 433 & 1 & 10k & \xval{f\_ht} & 438 e & 6,265 & 13,734k\\
\end{tabular}

\medskip

  \caption{Sequences $\DLEM$ generated via DAG compression. $D_0$ is composed
    from results of \SGCD runs \h{Base runs}, taking the components indicated
    in column $D_0$: M stands for the final $M_i$, A|$n$ for $n$ members from
    $\Abandoned$ (evenly distributed w.r.t. recording time). For $\LAL$, the
    50k refer to each base run, amounting to a total of 100k proof terms.
    Column \h{Values of}~\NLem lists prefix lengths with which the sequence
    was used in experiments. Further columns show cardinalities of
    intermediate results.}
  \label{tab-lemma-settings}
  \tablesettings  
  \centering 
  \setlength{\tabcolsep}{4pt}
  \begin{tabular}{llllrrr}
  $\DLEM$-\h{Id} & \h{Base runs} & $D_0$ & \h{Values of} \NLem & $|D_0|$ & $|\DLEM'|$ & $|\DLEM|$\\\midrule
$\LMS$ & $\RSBP,\RSBT$ &  M  & 500, 1k                      & 5,974   & 2,722  & 2,455\\
$\LMM$ & $\RMBP,\RMBT$ &  M  & 100                          & 22,870  & 10,830 & 9,877\\
$\LML$ & $\RLBP,\RLBT$ &  M  & 4k                           & 59,835  & 27,412 & 24,970\\
$\LMT$ & $\RLBT$     &  M  &  200                          & 9,998   & 3,840  & 3,806\\
$\LAL$ & $\RLBT,\RLBP$ & M, A|50k & 1k                     & 159,831 & 61,482 & 53,539 \\
$\LACP$ & $\RLBP$   & M, A|100k & 800, 1k, 1.5k, 12k     & 150,000 & 69,493 & 66,637\\
$\LACT$ & $\RLBT$   & M, A|100k & 100                    & 109,998 & 14,110 & 13,507
  \end{tabular}

  \vspace{-5pt}
\end{table}

Base sets $D_0$ in our experiments were composed from results of the six \SGCD
runs summarized in Table~\ref{tab-addaxioms-bases}. Configurations are as for
$\RBP$ and $\RBT$ (Table~\ref{tab-base-pt}) but with smaller \xpar{Trim} values.
$\RMBP$ and $\RLBP$ prove a few \POI theorems not in \POIBASE: a larger
\xpar{Trim} value does not imply that the result is a superset. Based on the
runs of Table~\ref{tab-addaxioms-bases} we synthesized different sequences
$\DLEM$, shown in Table~\ref{tab-lemma-settings}.

\xsubsection{Lemma-Enhanced Theorem Generation with \SGCD}
\label{sec-addaxioms-sgcd}

Table~\ref{tab-results-sav} summarizes runs of \SGCD with lemmas synthesized
by our workflow supplied as additional axioms. In all runs, \xpar{Rest} is set
to \xval{dup} and to \xval{gen\_max} with value \h{GMax}, \xpar{Ord} is
\xval{f\_th}, and (except for $\RS_{12}$) \xpar{Post} is \xval{subs}. Recall
that our virtual run $\Rbase$ defined the 554 \POIBASE theorems. Our
experiments with lemma synthesis yield proofs of 143 further \POI theorems.
Thus, in total, we now have generated proofs for 697 \POI theorems, including
10 with rating~1.

\begin{table}[t]
  \caption{Lemma-enhanced runs of \SGCD. $\DLEM$ refers to
    Table~\ref{tab-lemma-settings}. The first \NLem members of $\DLEM$ are
    taken as lemmas, i.e., their MGTs are given as axioms, along with the
    three axioms from Table~\ref{tab-propaxioms}. Column \h{Inp} shows the
    number of \POI theorems which are subsumed by an input lemma. The last
    four rows refer to unions of runs.}
  \label{tab-results-sav}
  \centering
  \tablesettings
  \setlength{\tabcolsep}{2.5pt}

\begin{tabular}{lcrrrrrcrrrrrr}
   \h{\RID} & $L$ & \h{NB} & \h{\POI} & \h{1} & \h{Inp} & \NLem & $\DLEM$ & \h{GMax} &  \xpar{Trim} & \h{Lev} & \h{Time} & \h{Mem} & $|\h{Gen}|$\\\midrule
$\RS_1$ & $\PSP$ & 61 & 556 & 1 & 146 & 500  & $\LMS$ & 40,000k & 80k & 7 & 2,990 & 62 & 36,882k\\
$\RS_2$ & $\PSP$ & 62 & 568 & 1 & 165 & 1,000 & $\LMS$ & 40,000k & 80k & 4 & 3,176 & 77 & 44,011k\\
$\RS_3$ & $\PSP$ & 15 & 503 & 3 & 62 & 100  & $\LMM$ & 40,000k & 80k & 78 & 3,518 & 20 & 11,556k\\
$\RS_4$ & $\PSP$ & 61 & 578 & 1 & 227 & 4,000 & $\LML$ & 40,000k & 80k & 2 & 3,117 & 117 & 46,219k\\
$\RS_5$ & $\PSP$ & 28 & 514 & 1 & 70 & 200  & $\LMT$ & 30,000k & 100k & 26 & 3,269 & 38 & 22,737k\\
$\RS_6$ & $\PSP$ & 43 & 558 & 5 & 70 & 200  & $\LMT$ & 30,000k & 200k & 7 & 2,901 & 52 & 31,848k\\
$\RS_7$ & $\PSP$ & 22 & 502 & 1 & 131 & 1,000 & $\LAL$ & 30,000k & 50k & 3 & 2,083 & 86 & 32,730k\\
$\RS_8$ & $\PSP$ & 36 & 549 & 2 & 127 & 800 & $\LACP$ & 30,000k & 100k & 3 & 2,453 & 82 & 35,688k\\
$\RS_9$ & $\HEIGHT$ & 42 & 533 & 1 & 137 & 1,000  & $\LACP$ & 40,000k & 80k & 2 & 2,866 & 124 & 40,559k\\
$\RS_{10}$ & $\HEIGHT$ & 64 & 574 & 2 & 137 & 1,000  & $\LACP$ & $\infty$ & 20k & 2 & 2,118 & 94 & 32,446k\\
$\RS_{11}$ & $\HEIGHT$ & 38 & 526 & 1 & 147 & 1,500  & $\LACP$ & 40,000k & 20k & 4 & 3,286 & 110 & 41,780k\\
$\RS_{12}$ & $\HEIGHT$ & 56 & 573 & 3 & 254 & 12,000 & $\LACP$ & 80,000k & $\infty$ & 1 & 1,186 & 93 & 80,012k\\
$\RS_{13}$ & $\PSP$ & 9  & 413 & 4  & 46 & 100  & $\LACT$ & $\infty$ & 100k & 4 & 803 & 19 & 10,285k\\\midrule
\multicolumn{2}{l}{$\RS_{1-6}$} & 115 & 655  & 5\\
\multicolumn{2}{l}{$\RS_{7-13}$} & 110 & 653 & 6\\
\multicolumn{2}{l}{$\RS_{1-13}$} & 143 & 692 & 8\\
\multicolumn{2}{l}{$\Rbase,\RS_{1-13}$} & 143 & 697 & 10\\ 
\end{tabular}

\bigskip

  \caption{Results of first-order provers on the 1,374 proving problems for
    the POI theorems (Sect.~\ref{sec-ideal}) with lemmas obtained by synthesis
    via proof term compression. Cells show the number of solved problems.
    \h{Base} refers to problems with just the three original axioms. The
    $\RS_i$ columns refer to settings from Table~\ref{tab-results-sav}: The
    axioms were enriched by the MGTs of the first \NLem proof terms from
    $\DLEM$. Best values are in bold.}
  \label{tab-lemmas-for-provers} 
  \centering
  \tablesettings  
  \setlength{\tabcolsep}{5pt}
  
  \begin{tabular}{lrrrrrrrr}
  \h{Prover} & \h{Base}   &  $\RS_{3}$  & $\RS_{13}$ & $\RS_{5}$ & $\RS_{2}$ & $\RS_{10}$ & $\RS_{4}$ & $\RS_{12}$\\\midrule
  \Vampire & 1,023        & \textbf{1,290}      & 1,284     & 1,250    &  1,180    & 1,179     & 1,091    & 1,064\\
  \EProver & 906   	  &   893      &   873     & 936      &   865     &  \textbf{999}      &   969    &   935\\
  \ProverN & 289   	  &   353      &   359     & 378      &   376     &  \textbf{438}      &   387    &   297\\
  \leanCoP & 90    	  &   527      &   410     & 505      &   557     &  607      &   601    &   \textbf{609}\\
\end{tabular} 
\vspace{-5pt}
\end{table}

We ``learned'' from sets of proofs of easy problems ($D_0$) how to prove
harder problems (``\h{NB} theorems'', i.e., theorems in \POI but not in
\POIBASE). The $D_0$ may then be viewed as training data, and the number of
proven \h{NB} theorems as evaluation criterion. Since the $D_0$ are computed
similarly as the proofs used to define \POIBASE, the $D_0$ contain (with 5
exceptions) no proofs of \h{NB} theorems.

The runs of Table~\ref{tab-results-sav} were chosen from 100 test runs such
that taken together they provide proofs for a large number of \h{NB} theorems.
The median of the number of proven \h{NB} theorems in all 100 runs is 31.5.
Runs not selected included those with fewer than 100 lemmas, and those with
$\TSIZE$ as level characterization. Nevertheless, $\TSIZE$ proved useful for
computing $D_0$. Also, both quickly obtained small sets as well as large sets
of abandoned proof terms from long runs proved useful for $D_0$.
Runs with many lemmas but low \xpar{MaxLevel} often give good results. Run
$\RS_{12}$ is extreme: with 12,000 lemmas it just generates level~1.

In the long term, lemma synthesis by compression should be integrated into
\SGCD. A naive approach fails due to an inherent mismatch with bottom-up level
generation: The save-value of a proof term $d$ is determined by proof terms
that have $d$ as strict subproofs, whereas for $\TSIZE$, $\PSP$, $\HEIGHT$ the
newly generated proof terms for $M'_i$ never are strict subproofs of the
available proofs in $M_{i-1}$ and $\Abandoned$.

\xsubsection{Lemma-Enhanced First-Order Theorem Proving}  
\label{sec-lemmas-fo-provers}

Table~\ref{tab-lemmas-for-provers} shows results of supplementing lemma sets
used for the \SGCD runs in Table~\ref{tab-results-sav} to first-order
provers.\footnote{Platform and provers as in footnote~\ref{foot-cluster}. Time
limit: 3,600~s. RAM limit: 3.77~GB.} The provers were not involved in
synthesis and selection of the lemmas. Significant improvements of the
solution rates suggest that lemma synthesis with proof term compression makes
a contribution that is orthogonal not just to top-down structure-generating
provers, such as \leanCoP, but also to the techniques in state-of-the-art
provers, such as \Vampire.

\xsubsection{Precision Aspects}
\label{sec-precision}

Our lemma synthesis provides an experimentally validated method to obtain sets
of useful lemmas, suggesting to inspect these also from the ``precision''
perspective, assessing the proportion of members in the ideal solution set.
As example, we consider the sequence $\LMM$ of proof terms specified in
Table~\ref{tab-lemma-settings}. Its members are already expected to have some
``value'', since they represent nodes with at least two incoming edges in the
minimal DAG for some underlying $D_0$. To get
\begin{wrapfigure}[12]{r}{4.7cm}
  \hspace{-6pt}
  \scalebox{0.575}{
    \begin{tikzpicture}
      \begin{semilogxaxis}[
      cycle list name=prec,      
      log basis x = {2},
      xmajorgrids = true,
      ymajorgrids = true,
      line width=1,
      xtick = {2,4,8,16,32,64,128,256,512,1024,2048},
      xlabel={Prefix length}, ylabel={Percentage of \POI proofs},      
    ]
        \addplot coordinates {(2,100)(4,75)(8,88)(16,75)(32,63)(64,50)(128,39)(256,27)(512,19)(1024,12)(2048,7)};
    \addplot coordinates {(2,100)(4,100)(8,88)(16,69)(32,63)(64,41)(128,41)(256,31)(512,20)(1024,11)(2048,7)};
    \addplot coordinates {(2,100)(4,100)(8,100)(16,88)(32,75)(64,55)(128,35)(256,30)(512,20)(1024,11)(2048,6)};
    \legend{{Save-value},{MGT tree size},{MGT height}}
      \end{semilogxaxis}
  \end{tikzpicture}}
  \vspace{-20pt}
  \caption{Percentage of prefix members that prove a \POI theorem.}

\label{fig-precision-main}
\end{wrapfigure}
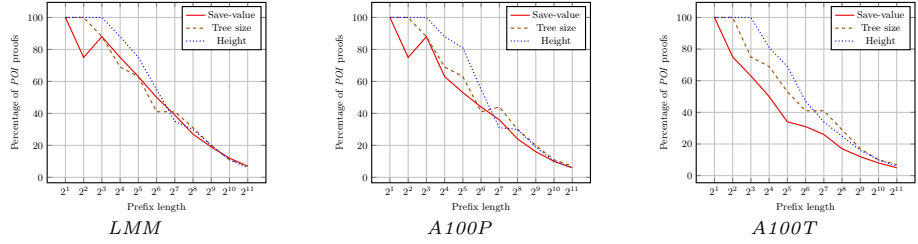
more fine-grained estimates, we
compare three re-orderings of the sequence: by save-value from synthesis, by
tree size of the MGT, and by height of the MGT. For prefixes of lengths
$2^1{-}2^{11}$, the percentage of members which prove a \POI theorem then
gives an idea of precision, as shown in Fig.~\ref{fig-precision-main}. E.g.,
in the prefix of length $2^6=64$ of $\LMM$ ordered by save-value $50\%$ (i.e.,
$32$) of the proof terms prove a \POI theorem. We see that MGT height gives
the best result up to a prefix length $2^6{-}2^7$, when MGT tree size and
save-value become competitive.

\section{Theorem Generation with Combinatory Proof Schemas}
\label{sec-combinatory}

In Sect.~\ref{subsec-compacted} we outlined the incorporation of combinators
into proof term construction \cite{cw:ccs:2022}. We show how this can be
utilized with generalizations of the PSP level and apply it for generating
theorems with \SGCD.

\xsubsection{Proof Patterns with Combinatory Definitions}

When enumerating proof terms, allowing arbitrary combinators in any position
can excessively increase the number of proof terms. Thus, we consider
mechanisms to allow only combinators from a given set, and only in given
contexts.

A \defname{proof pattern} $p/k$ is characterized by a function symbol~$p$ with
arity $k > 0$, and an associated combinator $c$, which determines semantics:
For proof terms $d_1,\ldots,d_k$, the MGT of $p(d_1,\ldots,d_k)$ is the MGT of
the corresponding instance $\D(\D(\D(c,d_1),d_2), \ldots, d_k)$ of the
pattern's \emph{combinatory definition}, taking as MGT of $c$ its principal
type. Proof terms are then formed from proof pattern functions instead of
$\D$. By rewriting patterns with their combinatory definition, they can be
converted to proof terms with $\D$ as only function symbol and combinators in
axiom position. E.g., let $\Patterns = \{\fp_1/2, \fp_2/2\}$, where $\fp_1$ is
associated with $\c{B}$, and $\fp_2$ with $\c{I}$. The following DAG grammar
then represents a proof term.
\begin{equation}
  \label{eq-proof-pattern}
\{4 \pro \fp_1(1, 1),\; \mathit{Start} \pro \fp_2(4, \fp_2(4, \fp_2(4, 1)))\}
\end{equation}
By rewriting $\fp_1(V_1,V_2)$ with $\D(\D(\c{B},V_1),V_2)$, the combinatory
definition of $\fp_1$, and $\fp_2(V_1,V_2)$ with $\D(V_1,V_2)$, we obtain
(\ref{eq-indu-d2-dag}) in Sect.~\ref{subsec-compacted}. For rewriting $\fp_2$
we used $\D(V_1,V_2)$, obtained by combinator reduction from
the longer combinatory definition
$\D(\D(\c{I},V_1),V_2)$.

To convert a proof term with combinators to one with only original axioms,
combinators can be rewritten with their reduction rules, as shown with
(\ref{eq-indu-d2-expanded}), which may lead to exponential increase of
compacted size. A possibility to eliminate combinators with linear increase
opens up if their principal types can be proven from original axioms:
occurrences of combinators in proof terms can then be replaced by proofs of
their principal type. This holds for our axioms from
Table~\ref{tab-propaxioms}, where proofs of $\c{K}$ and $\c{S}$ are directly
available as axioms~1 and~2. E.g, the principal type of $\c{B}$ is $(x\oimp
y)\oimp((z\oimp x)\oimp(z\oimp y))$ and is the MGT of $\D(\D(2, \D(1, 2)),
1)$. Or, in other words, $\c{B}$ can be defined as $\c{S} (\c{K} \c{S})
\c{K}$.
Hence, the production $4 \pro \D(\D(\c{B}, 1), 1)$ in (\ref{eq-indu-d2-dag})
can be converted to $4 \pro \D(\D(\D(\D(2, \D(1, 2)), 1), 1), 1)$.
For a larger example see \appref{appsec-pattern-proofs}.

A third possibility for linearly eliminating combinators is by conversion to a
tree grammar \cite{cw:zz:2025,lohrey:survey:2015,lohrey:treerepair:2013}. We
do not pursue this here but note an observation: Since conversion of a tree
grammar to a proof with combinators is essentially $\lambda$ to SKI
conversion, which can be performed in linear time and with linear size
increase \cite{kiselyov:lambdaToSKI:2018}, it follows that if the principal
types of the combinators can be proven from the original axioms, then the size
of a tree grammar compression can improve at most linearly over the size of
the minimal DAG compression.

\xsubsection{The PSP Level for Proof Schemas}
\label{sec-patterns-intro}
We assume a finite set $\Ax$ of axiom identifiers. For proof terms built from
proof patterns, we generalize the inductive characterization of the PSP level
as follows.

\begin{defn}[The PSP Level for Proof Patterns]
  Assume a finite set $\Patterns$ of proof patterns. Define the \name{PSP
    level for proof patterns} as
  {%
  \[\begin{array}{lcl}
  \PP_0 & \eqdef & \Ax;\\
  \PP_{n+1} & \eqdef & \{p(d_1,\ldots,d_{i-1},d_i,d_{i+1}\ldots,d_k)
  \; \mid\\
  && \hparen p/k \in \Patterns,\; i \in \{1,\ldots,k\},\; d_i \in \PP_n,\\
  && \hparen \mathit{for }\; j \in \{1,\ldots,k\} \setminus \{i\}\; .\; d_j \in \ST(d_i) \cup \Ax\}.
  \end{array}\]}
\end{defn}
We also consider a second generalization, where, in addition to proof
patterns, \name{direct proof schemas}, proof terms $\D(\D(\D(c,V_1),V_2),
\ldots, V_k)$ representing combinatory definitions of patterns can be
constituents of proof terms. A direct proof schema $c/k$ is characterized by a
combinator $c$ and an arity $k > 0$. Direct subterms of a pattern instance
$p(d_1,\ldots,d_k)$ are the arguments $d_1,\ldots,d_k$, while for a direct
schema instance $\D(\D(\D(c,d_1),d_2), \ldots, d_k)$ also $c$ itself and
applications of $c$ to any prefix of the arguments, i.e., terms
$\D(\D(\D(c,d_1),d_2), \ldots, d_i)$ for $i < k$, are strict subterms.

\begin{defn}[Characterization by PSP Level with Proof Schemas]
  Assume a finite set $\Schemas$ of proof patterns and direct proof schemas.
  Define the \name{PSP level with proof schemas} as
  \setlength{\arraycolsep}{1pt}
  {%
  \[\begin{array}{lcl}
  \PK_0 & \eqdef & \Ax;\\
  \PK_{n+1} & \eqdef &  \hphantom{\cup\;} \{\D(d,e) \mid d \in \PK_n,\; e \in \ST(d) \cup \Ax \}\\
  && \cup\; \{\D(d,e) \mid e \in \PK_n,\; d \in \ST(e) \cup \Ax \}\\
  && \cup\; \{t[d_1,\ldots,d_k] \mid t/k \in \Schemas,
  \mathit{for }\; i \in \{1,\ldots,k\}\; .\; d_i
  \in \PK_n \cup \Ax\},\end{array}\]}where if $t/k$ is a proof pattern $p/k$,
  then $t[d_1,\ldots,d_k] = p(d_1,\ldots,d_k)$, and if $t/k$ is a direct
  schema $c/k$, then $t[d_1,\ldots,d_k] = \D(\D(\D(c,d_1),d_2), \ldots,
  d_k)$.
\end{defn}

\enlargethispage{0.3cm}
\xsubsection{Theorem Generation with \SGCD and Proof Schemas}

\begin{table}[t]
  \caption{Runs of \SGCD with level characterization $\PP$ for the specified
    patterns. The last three rows refer to unions of runs.}
  \label{tab-results-pp}    
  \centering
  \tablesettings  
\setlength{\tabcolsep}{3.1pt}
\begin{tabular}{llrrrlrrrrrr}
  \h{\RID} & $L$  & \h{NB} & \h{\POI} & \h{1} & \h{Patterns} & \hspace{-4pt} \h{GMax} & \xpar{Trim} & \h{Lev} & \h{Time} & \h{Mem} & \h{Gen}\\\midrule
$\RC_1$ & $\PP$ & 19 & 379 & 1 & $\c{I}/2,\c{B_{4}},\c{C},\c{C_{4}},\c{S_{4}}$ & 133k & 13k & 20 & 540 & 7.00 & 2,280k\\
$\RC_2$ & $\PP$ & 15 & 363 & 5 & $\c{I}/2,\c{B_{4}},\c{C},\c{S}$ & 67k & 10k & 51 & 3,473 & 56.08 & 2,508k\\
$\RC_3$ & $\PP$ & 14 & 397 & 2 & $\c{I}/2,\c{B_{4}},\c{C_{4}},\c{B}$ & 133k & 13k & 27 & 1,166 & 14.30 & 2,976k\\
$\RC_4$ & $\PP$ & 13 & 403 & 2 & $\c{I}/2,\c{B_{4}},\c{C},\c{B},\c{S_{4}}$ & 133k & 13k & 48 & 1,609 & 31.96 & 2,651k\\
$\RC_5$ & $\PP$ & 12 & 407 & 3 & $\c{I}/2,\c{B_{4}},\c{C},\c{S}$ & 200k & 20k & 25 & 1,261 & 17.13 & 4,092k\\
$\RC_6$ & $\PP$ & 10 & 397 & 3 & $\c{I}/2,\c{B_{4}},\c{C},\c{C_{4}}$ & 533k & 53k & 14 & 1,162 & 15.62 & 5,783k\\
$\RC_7$ & $\PP$ & 9 & 383 & 0 & $\c{I}/2,\c{C_{4}},\c{S}$ & 133k & 13k & 20 & 603 & 8.96 & 2,234k\\
$\RC_8$ & $\PP$ & 5 & 301 & 0 & $\c{I}/2,\c{C},\c{S}$ & 200k & 20k & 20 & 317 & 3.16 & 1,453k\\
$\RC_9$ & $\PP$ & 2 & 320 & 0 & $\c{I}/2,\c{B_{4}} (\c{C_{4}} \c{C})$ & 80k & 8k & 17 & 937 & 20.24 & 1,127k\\
$\RC_{10}$ & $\PP$ & 26 & 213 & 0 & $\c{B_{4}},\c{C_{4}}$ & 267k & 27k & 21 & 1,348 & 19.38 & 4,462k\\
$\RC_{11}$ & $\PP$ & 14 & 228 & 1 & $\c{B_{4}},\c{C_{4}}$ & 400k & 40k & 18 & 1,588 & 19.59 & 5,916k\\
$\RC_{12}$ & $\PP$ & 12 & 182 & 0 & $\c{B},\c{C}$ & 200k & 20k & 20 & 261 & 2.78 & 1,319k\\
$\RC_{13}$ & $\PP$ & 11 & 362 & 1 & $\c{B},\c{B_{4}},\c{C},\c{C_{4}} \c{C}$ & 133k & 13k & 19 & 513 & 7.83 & 2,125k\\
$\RC_{14}$ & $\PP$ & 4 & 197 & 0 & $\c{B},\c{C_{4}},\c{S},\c{S_{4}}$ & 133k & 13k & 20 & 417 & 4.68 & 1,940k\\
$\RC_{15}$ & $\PP$ & 3 & 263 & 0 & $\c{B_{4}} \c{B}$ & 100k & 10k & 12 & 185 & 3.12 & 882k\\\midrule
\multicolumn{2}{l}{$\RC_{1-9}$} & 50 & 525 & 6\\
\multicolumn{2}{l}{$\RC_{10-15}$} & 46 & 443 & 1\\
\multicolumn{2}{l}{$\RC_{1-15}$} & 85 & 567 & 6\\
\end{tabular}

\medskip

  \caption{Runs of \SGCD with level characterization $\PK$ for the specified
    schemas. The last four rows refer to unions of runs, now also
    considering $\RS_{1-13}$ from Table~\ref{tab-results-sav}.}
  \label{tab-results-pk}  
  \centering
  \tablesettings  
\setlength{\tabcolsep}{1.3pt}  
\begin{tabular}{llrrrlrrrrrr}
  \h{\RID} & $L$  & \h{NB} & \h{\POI} & \h{1} & \h{Schemas} & \hspace{-6pt}\h{GMax} & \xpar{Trim} & \h{Lev} & \h{Time} & \h{Mem} & \h{Gen}\\\midrule  
$\RC_{16}$ & $\PK$ & 6 & 209 & 0 & $\c{B},\c{C},\c{S},\c{S_{4}}$ & 200k & 20k & 9 & 950 & 12.9 & 1,366k\\
$\RC_{17}$ & $\PK$ & 5 & 190 & 0 & $\c{B},\c{B_{4}},\c{S}$ & 133k & 13k & 12 & 2,716 & 50.42 & 1,305k\\
$\RC_{18}$ & $\PK$ & 17 & 223 & 0 & \DIRECT $\c{B},\c{B_{4}},\c{C_{4}},\c{S}$ & 133k & 13k & 10 & 2,632 & 45.59 & 1,066k\\
$\RC_{19}$ & $\PK$ & 6 & 235 & 0 & \DIRECT $\c{B},\c{S_{4}}$ & 44k & 4k & 12 & 1,482 & 28.29 & 403k\\
$\RC_{20}$ & $\PK$ & 12 & 413 & 1 & \DIRECT $\c{B}/1,\c{C}/1,\c{C_{4}}/1,\c{S_{4}}/1$ & 400k & 40k & 20 & 341 & 3.47 & 1,685k\\
$\RC_{21}$ & $\PK$ & 9 & 332 & 1 & \DIRECT $\c{C_{4}} (\c{C_{4}} (\c{C_{4}} (\c{C_{4}} \c{C})))/1$ & 400k & 40k & 20 & 782 & 7.30 & 3,146k\\
$\RC_{22}$ & $\PK$ & 7 & 360 & 0 & \DIRECT $\c{B_{4}} (\c{C_{4}} \c{C})/1$ & 400k & 40k & 20 & 515 & 4.91 & 2,228k\\\midrule
\multicolumn{2}{l}{$\RC_{16-22}$} & 45 & 503 & 1\\  
\multicolumn{2}{l}{$\RC_{1-22}$} & 116 & 633 & 6\\
\multicolumn{2}{l}{$\Rbase,\RC_{1-22}$} & 116 & 670 & 10\\
\multicolumn{2}{l}{$\Rbase,\RS_{1-13},\RC_{1-22}$} & 180 & 734 & 12\\
\end{tabular}
\vspace{-11pt}
\end{table}

To get initial sets of combinators that seem useful as proof schemas we
analyzed \SETMM \apprefparen{app-finding-combs}. Following
\cite{peytonjones:87}, we considered as primitive combinators $\c{S}, \c{K},
\c{I}, \c{B}, \c{C}, \c{S_4},$ $\c{B_4}, \c{C_4}$. The latter (called
$\c{S'}, \c{B^*}, \c{C'}$ in \cite{peytonjones:87}) are defined as follows.
{\tablesettings
\[\begin{array}{l@{\hspace{4pt}}l@{\hspace{4pt}}l}
        & \text{\textit{$\lambda$-Term}} & \text{\textit{Principal Type}} \\\midrule
\c{S_4} &  \lambda V_1V_2V_3V_4\; .\\   
& \D(\D(V_1, \D(V_2, V_4)), \D(V_3, V_4)) 
  & (x\oimp (y\oimp z))\oimp ((u\oimp x)\oimp ((u\oimp y)\oimp (u\oimp z)))\\
  \c{B_4} &  \lambda V_1V_2V_3V_4 . \D(V_1, \D(V_2, \D(V_3, V_4)))
  & (x\oimp y)\oimp ((z\oimp x)\oimp ((u\oimp z)\oimp (u\oimp y)))\\
  \c{C_4} & \lambda V_1V_2V_3V_4 . \D(\D(V_1, \D(V_2, V_4)), V_3)
  & (x\oimp (y\oimp z))\oimp ((u\oimp x)\oimp (y\oimp (u\oimp z)))\\  
\end{array}\]}

Table~\ref{tab-results-pp} summarizes \SGCD runs with level characterization
$\PP$. Members of \textit{Patterns} are specified by their defining
combinator. As arity, by default the standard arity of the combinator minus
one is taken, where the \defname{standard arity} is the minimal $k$ such that
application of the combinator to $k$ arguments reduces to a term without the
combinator. E.g., the standard arity of $\c{B}$ and $\c{C}$ is 3, and that of
$\c{B_4}$, $\c{C_4}$ and $\c{S_4}$ is 4. Hence, e.g., the default arity for a
pattern with defining combinator $\c{B}$ is 2. The standard arity of $\c{I}$
is 1. We use $\c{I}$ as pattern with arity 2 to express $\D$, as in
(\ref{eq-proof-pattern}) above. Both, runs with $\c{I}/2$ and without it,
yield proofs of largely disjoint sets of \POI theorems not in \POIBASE, 85 in
total.

Table~\ref{tab-results-pk} summarizes \SGCD runs with level characterization
$\PK$. Members of \textit{Schemas} are patterns as specified for
Table~\ref{tab-results-pp}, or, indicated by \textit{d}, direct proof schemas,
specified analogously by the combinator.
In runs of Tables~\ref{tab-results-pp} and~\ref{tab-results-pk} \xpar{Rest} is
set to \xval{dup} and to \xval{gen\_max} with value \h{GMax}, \xpar{Ord} is
\xval{f\_th}, and \xpar{Post} is \xval{subs}. As in the experiments described
in Sect.~\ref{sec-addaxioms-sgcd}, runs were chosen from a larger test set
such that taken together they prove a large number of \POI theorems not in
\POIBASE.
Runs with $\PK$ yield proofs of further such theorems, such that in total the
runs with combinatory proof schemas contribute proofs for 116 \POI theorems
not in \POIBASE. Together with the results of lemma synthesis from
Sect.~\ref{sec-addaxioms-sgcd}, we finally have generated proofs for 734 of
the 1,374 \POI theorems, including 12 with rating~1. Those with rating 1 are
re-inspected in \appref{app-hard-problems}. Details for each \POI theorem are
given in \appref{app-bigtable}.

\section{Conclusion}
\label{sec-conclusion}

To study generating theorems from given axioms, we extracted 1,374 theorems
from \SETMM as a benchmark, and developed techniques that could generate 734
of these. Our synthesized lemmas significantly improved solution rates of
automated provers on proving problems for the benchmark theorems.
Our approach is centered on proof structures as terms, which are enumerated
and compressed, differently from most works with similar aims, which are
centered on formulas, e.g.,
\cite{wos:resonance:95,vyskocil:stanovsky:urban:definitions:2010,hetzl:tree:2012,lemmanaid:2025,axelrod:aitp:2025}.
An early exception is work by Stephan Schulz
\cite{schulz:projektarbeit:1993,denzinger:schulz:1994}, where, essentially,
save-values in a proof DAG identify important steps. This and other measures
based on proof graphs were later investigated with large \name{HOL light}
corpora for lemma synthesis and premise selection
\cite{kaliszyk:urban:millions:2015}.

Our uses of proof term compression call for further studies of DAGs vs. tree
grammars, detecting combinatory schemas in \SETMM, and investigating specific
combinatory schemas.
With \name{inductive level characterizations} we presented an abstract
framework and a family of practical proving techniques, reimagining proving by
proof enumeration, which seems worth further investigation.
Follow-up tasks with the \SGCD system include iteration and combination of
presented techniques, automated parameter calibration, and experiments with it
as prover, utilizing the interplay of top-down and bottom-up processing.

Although, so far, our experiments were limited to the fundamentals of \SETMM,
its axioms for propositional logic, our techniques and implementation are
ready for other axioms, arbitrary first-order Horn formulas.
Predicate logic in \SETMM would require more: a second function in proof terms
\cite{megill:1995}, and axioms that seem to benefit from adapting equality
handling of saturation-based provers.

Essential for our approach are proof terms with a most general theorem. For
resolution, this seems possible if proofs are constrained
\cite{lee:thesis:1967} or transformed \cite[Lemma~4.1.3]{kleinebuening:1999}
such that final steps are for goals. Another path to resolution would be
simulation by combinators in proof terms \cite{cw:ccs:2022}, related to
\cite{eder:cs:1989}.

For automated proving that involves machine learning with proof structures
\cite{rwzb:lemmas:2023,hindsight:2024,cw:zz:ml:2026}, our work provides a
foundation, an implemented infrastructure, and a yardstick for comparison with
capabilities of symbolic techniques.

\clearpage

\subsection*{Acknowledgments}
The author would like to thank Zsolt Zombori for the fruitful exchange during
work on this project, Tim Richter for technical support, and anonymous
reviewers for helpful suggestions to improve the presentation.

Funded by the Deutsche Forschungsgemeinschaft (DFG, German Research
Foundation) -- Project-ID~457292495. The author gratefully acknowledges the
computing time granted by the Resource Allocation Board and provided on the
supercomputer Emmy/Grete at NHR-Nord@Göttingen as part of the NHR
infrastructure. The calculations for this research were conducted with
computing resources under the project
\texttt{nhr\_bb\_test\_proof\_structures}.

\bibliographystyle{splncs04}
\bibliography{bibthgen01}

\clearpage
\appendix

\renewcommand{\contentsname}{Appendices}
\setcounter{tocdepth}{2}
\tableofcontents

\clearpage
\resumetoc

\section{Supplementary Material for Section~\ref{sec-ideal}: Conversion to Implication Form}
\label{app-implication-form}

\subsubsection{A Tree-Grammar View of \SETMM.}

The \MM database \SETMM builds up mathematics with an inference system that
has two primitive proof term constructors: the binary condensed detachment
$\D$ to handle propositional inferences, and the unary condensed
generalization $\G$ to handle quantification.

A theorem formula of \SETMM can be technically viewed as a definite
first-order clause with the single unary predicate $\P$. Generalizing the
specification in Sect.~\ref{sec-bg}, proof terms can now have variables $V_1,
V_2, \ldots$, which we call \emph{parameters} to distinguish them form formula
variables. The MGT of a proof term with parameters is a definite clause with a
body atom for each parameter, the $i$-th body atom corresponding to $V_i$. The
handling of free variables in determining the MGT is specified in
\cite{cw:zz:2025}.\footnote{It is very similar to the consideration of a
\emph{context} in computing the principal type \cite{hindley:book:1997}. For
the definite clause MGT, however, the correspondence between parameters and
positions in the clause body has to be obeyed.}

A tree grammar \cite{lohrey:survey:2015,lohrey:treerepair:2013} that is
non-cyclic and has a single production for each nonterminal provides a
compressed representation of a set of proof terms \cite{cw:zz:2025}. It
generalizes DAG grammars in that RHSs can now have parameters, which appear as
arguments of the nonterminals forming the LHSs.
Accordingly, the set of all proofs in the \SETMM knowledge base can be viewed
as a single tree grammar with a production for each theorem. This tree grammar
describes in compressed form a set of proof trees with only $\D$ and $\G$ as
labels of inner nodes and axiom identifiers as labels of leaves. The members
of the represented set are proofs of the ``top-level'' theorems, i.e.,
theorems that are not referenced in the proof of another theorem. Nonterminals
then appear, aside of $\D$, $\G$ and axiom identifiers, as function symbols in
the proof terms on the RHSs of the grammar.

As an example consider theorem \mm{mpi}:
\[\P(y\oimp z) \revimp (\P(x) \land \P(y\oimp (x\oimp z))).\]
A direct translation of its proof in \SETMM leads to the following tree
grammar.
The proof makes use of theorems \mm{mpd}, \mm{a2i} and \mm{a1i} as lemmas. As
axioms we take those from Table~\ref{tab-propaxioms}.
\[
\begin{array}{lcl}
 \mm{a1i}(V_1)&\pro& \D(1, V_1)\\
 \mm{a2i}(V_1)&\pro& \D(2, V_1)\\
 \mm{mpd}(V_1, V_2)&\pro& \D(\mm{a2i}(V_2), V_1)\\
 \mm{mpi}(V_1, V_2)&\pro& \mm{mpd}(\mm{a1i}(V_1), V_2).\\
\end{array}
\]
The grammar describes the following tree.
\[
\D(\D(2, V_2), \D(1, V_1))).
\]

\enlargethispage{2mm}
\subsubsection{Conversion of Definite Clauses to Implication Form.}

From a first-order point of view, \MM theorems are definite clauses $\P(g)
\revimp (\P(f_1) \land \ldots \land \P(f_n))$
\cite{cw:zz:2025,metamath:atp:2023}. We convert them to unit clauses $\P(f_1
\oimp (f_2 \oimp \ldots (f_n \oimp g)))$, which we call \emph{in implication
form}. Thus, in our first-order view we have the detachment clause
(\ref{eq-det}) and a clause for the second primitive, condensed
generalization, as the sole non-unit clauses. For example, theorem \mm{mpi}
would be converted to
\[\P(x\oimp ((y\oimp (x\oimp z))\oimp (y\oimp z))).\]
This conversion has no essential impact as proofs for both forms can be
linearly translated into each other, which can be justified as follows.

From a proof $d$ of the unit form, we can easily obtain a proof $d'$ of the
corresponding definite clause as follows. If $d$ has the MGT $\P(f_1 \oimp
(f_2 \oimp \ldots (f_n \oimp g) \ldots ))$, then $d' = \D(\ldots \D(\D(d,
V_1), V_2) \ldots, V_n)$ has as MGT the definite clause $\P(g) \revimp
(\P(f_1) \land \ldots \land \P(f_n))$. The parameters of the proof term $d'$
are $V_1, \ldots, V_n$. For example, if $d=1$, then $d$ has the MGT $\P(x
\oimp (y \oimp x))$, and $d' = \D(\D(1, V_1),V_2)$ has as MGT the definite
clause $P(x) \revimp (P(x) \land \P(y))$.

Vice versa, from a proof $d'$ with parameters $V_1,\ldots,V_n$ (whose MGT is a
definite clause with body length $n$ such that the $i$-th body atom
corresponds to $V_i$) a proof $d$ with the corresponding unit form as MGT can
be obtained in the following steps. (1) Eliminate the variables from $d'$ by
compiling $\lambda V_1 . \ldots \lambda V_n . d'$ to a SKI combinator
\cite{peytonjones:87,kiselyov:lambdaToSKI:2018}, reading the detachment
operator $\D$ as application. (2)~In the SKI combinator, replace occurrences
of primitive combinators by proofs of their principal type. Actually, since
principal type of $\c{K}$ is the MGT of axiom 1 and the principal type of
$\c{S}$ is the MGT of axiom 2, this amounts to replacing the primitive
combinators by characterizations in terms of $\c{K}$ and $\c{S}$, followed by
replacing $\c{K}$ with $1$ and $\c{S}$ with 2. For example, if $d' =
\D(\D(1,V_2),V_1)$, its MGT is $\P(y) \revimp (\P(x) \land \P(y))$ (recall
that $V_i$ corresponds to the $i$-th body atom). Then $\lambda V_1 . \lambda
V_2 . \D(\D(1,V_2),V_1)$ compiles into $\D(\c{C},1)$. In the notation of
combinatory logic $\c{C}$ can be characterized as $\c{S} (\c{K} (\c{S} \c{S}
(\c{K} \c{K}))) (\c{S} (\c{K} \c{K}) \c{S})$. Accordingly, in step (2) $\c{C}$
is replaced with $\D(\D(2, \D(1, \D(\D(2, 2), \D(1, 1)))), \D(\D(2, \D(1, 1)),
2))$, whose MGT is $((x\oimp (y\oimp z))\oimp (y\oimp (x\oimp z)))$, the
principal type of $\c{C}$.
Although the proof term may be enlarged by this conversion from a proof of a
definite clause to a proof of the corresponding unit, the conversion effort as
well as the size increase can actually be kept linear, due to a linear
$\lambda$-to-SKI compilation \cite{kiselyov:lambdaToSKI:2018}.

\clearpage
\section{Supplementary Material for
  Section~\ref{sec-generating-proof-structures}: Cardinalities of Levels}
\label{app-generating-proof-structures}

For level characterizations $L$, we can compare the sequence of the
cardinalities of the levels $L_i$. It gives an impression about the levels
that can be reached easily in practice and the required extent of reductions
to reach higher levels. For a single axiom $|\TSIZE_i|$ increases according to
\OEISNUM{A000108}\footnote{oeis:$N$ refers to sequence $N$ in the \name{OEIS}
\cite{oeis}.} (Catalan numbers), $|\CSIZE_i|$ according to \OEISNUM{A001699},
and $|\HEIGHT_i|$ according to \OEISNUM{A254789} \cite{cwwb:article:2024}. For
three axioms, cardinalities of initial levels are shown in
Table~\ref{tab-cards-full}. With three axioms, we find an \name{OEIS} entry
only for $|T_i|$ with \OEISNUM{A025226}.

\begin{table}[h]
  \centering
  \caption{Cardinalities of initial levels for three axioms. For level
    characterizations $L \in \{\TSIZE, \PSP, \CSIZE, \HEIGHT\}$, column
    $|L_i|$ shows the cardinality of $L_i$, column $|L_i{\cap}\MGT|$ the
    number of proof terms in $L_i$ that have a defined MGT for the axioms of
    Table~\ref{tab-propaxioms}, and column $|\MGT(L_i)|$ the number of
    distinct MGTs of the proof terms in $L_i$, for these axioms.}
  \label{tab-cards-full}
  \small
  \setlength{\arraycolsep}{4pt} $\begin{array}{rrrrrrrr} i & |\TSIZE_i| &
    |\TSIZE_i{\cap}\MGT| & |\MGT(\TSIZE_i)| & |\PSP_i| & |\PSP_i{\cap}\MGT| &
    |\MGT(\PSP_i)|\\\midrule
  0 & 3         & 3       & 3     & 3          & 3         & 3     \\
  1 & 9         & 6       & 6     & 9          & 6         & 6     \\
  2 & 54        & 24      & 18    & 63         & 28        & 19    \\
  3 & 405       & 132     & 60    & 567        & 180       & 60    \\
  4 & 3,402     & 729     & 166   & 6,237      & 1,460     & 193   \\
  5 & 30,618    & 4,628   & 516   & 81,081     & 14,348    & 682   \\
  6 & 288,684   & 29,541  & 1,665 & 1,216,215  & 169,413   & 2,597 \\
  7 & 2,814,669 & 198,617 & 5,282 & 20,675,655 & 2,329,830 & 11,096\\
\end{array}$

\medskip

$\begin{array}{rrrrrrrr}
  i & |\CSIZE_i| & |\CSIZE_i{\cap}\MGT| & |\MGT(\CSIZE_i)| & |\HEIGHT_i| & |\HEIGHT_i\cap\MGT| & |\MGT(\HEIGHT_i)|\\\midrule
  0 & 3          & 3         & 3      & 3           & 3         & 3    \\
  1 & 9          & 6         & 6      & 9           & 6         & 6    \\
  2 & 63         & 28        & 19     & 135         & 57        & 30   \\
  3 & 639        & 209       & 71     & 21,465      & 2,693     & 247  \\
  4 & 8,415      & 1,950     & 256    & 467,056,935 & 5,466,532 & 6,813\\
  5 & 136,431    & 23,548    & 1,020  \\
  6 & 2,634,687  & 347,395   & 4,456  \\
  7 & 59,204,223 & 6,153,424 & 21,315 \\
\end{array}$
\end{table}

\FloatBarrier
\clearpage
\section{Supplementary Material for Section~\ref{sec-basesolved}: Comparison of Proof Sizes}
\label{app-proofsize}

\ProverN in \code{auto} mode selects positive hyperresolution for CD problems,
which permits direct translation to our proof terms.
Table~\ref{tab-proof-sizes} compares proof sizes for those 299 theorems for
which proofs were obtained from \ProverN in a single cluster run over the TPTP
problems and \SGCD in the runs $\RBP$ and $\RBT$. For problems with several
proofs by \SGCD the smallest, compared lexically by compacted size, tree size
and height, was picked.
\begin{table}
  \caption{Size of proofs in comparison. For the multisets of 299 proof terms,
    their compacted size and their tree size are represented by median,
    average and maximum values. The minimum is in all cases 2. Column \h{all}
    shows the compacted size of the whole set of the 299 proof terms.}
\label{tab-proof-sizes}
\centering
\setlength{\tabcolsep}{4pt}
\begin{tabular}{lrrrrrrr}
  & \multicolumn{4}{c}{\h{Compacted size}} & \multicolumn{3}{c}{\h{Tree size}}\\
  \cmidrule(lr){2-5}\cmidrule(lr){6-8}
  \h{Source} & \h{med} & \h{avg} & \h{max} & \h{all} & \h{med} & \h{avg} & \h{max}\\\midrule
  \SGCD $\RBP,\RBT$ & 14 & 15.38 & 37 & 737 & 22 & 77.22 & 1,336\\
  \ProverN & 26 & 24.80 & 71 & 464 & 63.00 & 89.29 & 801\\
  \SETMM & 159 & 165.39 & 463 & 5,249 & 2491.00 & 3,239,638.77 & 461,218,315\\
\end{tabular}
\end{table}

Interestingly, the value of \h{all} is for \ProverN lower than for \SGCD,
although with \ProverN each proof term was obtained from a separate prover
invocation.

The comparison also shows values for proofs from \SETMM, expanded to full
proof terms built just from $\D$ and the three axiom identifiers, which
involved the proof transformations indicated in Sect.~\ref{sec-ideal} and
\appref{app-implication-form}. The \MM user actually does not work with DAG
compression but with a stronger grammar compression, and also with additional
definitional axioms. The proofs of the 299 original theorems are described in
\SETMM by a tree grammar with 567 productions
and grammar size (based on the number of edges) 1,667. The compacted size of
the set of all proof terms shown in column \h{all} has to be doubled for
comparison, i.e., considered as special case of a grammar, the DAG compression
of the set of the transformed proofs from \SETMM has grammar size 10,498.

\clearpage
\section{Supplementary Material for Section~\ref{sec-dag-synthesis}: Significance
of the Save-Value}
\label{app-supp-dag-synthesis}

To gain evidence of the actual significance of the save-value for lemma
selection, a criterion entirely in terms of proof-structures in contrast to
formulas, we repeated the experiments from Table~\ref{tab-results-sav} with an
altered ordering of the productions of grammar $G$ during synthesis: We now
compared \emph{first by the tree size} of the MGT, second by the height of the
MGT, and third by the save-value. The results, shown in
Table~\ref{tab-results-fth}, then were significantly worse, with only 94
instead of 143 proofs for \POI theorems not in \POIBASE.

\begin{table}[h]
  \caption{Re-runs of the experiments from Table~\ref{tab-results-sav} with
    ordering of productions in lemma synthesis primarily by MGT size measures
    instead of save-value.}
  \label{tab-results-fth}
  \centering 
  \setlength{\tabcolsep}{2.5pt}

\begin{tabular}{lcrrrrcrrrrrr}
   \h{\RID} & $L$ & \h{NB} & \h{\POI} & \h{1} & \NLem & $\DLEM$ & \h{GMax} &  \xpar{Trim} & \h{Lev} & \h{Time} & \h{Mem} & $|\h{Gen}|$\\\midrule
$\RS'_{1}$ & $\PSP$ & 34 & 484 & 1 & 500 & $\LMS$ & 40,000k & 80k & 7 \hse& 1,612 & 33 & 21,910k\\
$\RS'_{2}$ & $\PSP$ & 48 & 477 & 1 & 1,000 & $\LMS$ & 40,000k & 80k & 4 \hse& 2,671 & 62 & 38,190k\\
$\RS'_{3}$ & $\PSP$ & 30 & 490 & 3 & 100 & $\LMM$ & 40,000k & 80k & 37 e& 1,336 & 9 & 5,703k\\
$\RS'_{4}$ & $\PSP$ & 51 & 553 & 1 & 4,000 & $\LML$ & 40,000k & 80k & 2 \hse& 2,915 & 114 & 43,239k\\
$\RS'_{5}$ & $\PSP$ & 37 & 502 & 1 & 200 & $\LMT$ & 30,000k & 100k & 26 \hse& 2,048 & 23 & 14,368k\\
$\RS'_{6}$ & $\PSP$ & 40 & 535 & 3 & 200 & $\LMT$ & 30,000k & 200k & 7 \hse& 2,299 & 34 & 23,059k\\
$\RS'_{7}$ & $\PSP$ & 33 & 432 & 0 & 1,000 & $\LAL$ & 30,000k & 50k & 3 \hse& 1,094 & 40 & 17,502k\\
$\RS'_{8}$ & $\PSP$ & 38 & 438 & 0 & 800 & $\LACP$ & 30,000k & 100k & 3 \hse& 1,999 & 56 & 28,483k\\
$\RS'_{9}$ & $\PSP$ & 25 & 411 & 0 & 1,000 & $\LACP$ & 40,000k & 80k & 2 \hse& 2,879 & 123 & 40,204k\\
$\RS'_{10}$ & $\PSP$ & 37 & 447 & 0 & 1,000 & $\LACP$ & $\infty$ & 20k & 2 \hse& 2,448 & 103 & 33,511k\\
$\RS'_{11}$ & $\PSP$ & 58 & 530 & 1 & 1,500 & $\LACP$ & 40,000k & 20k & 4 \hse& 3,119 & 101 & 43,252k\\
$\RS'_{12}$ & $\PSP$ & 58 & 557 & 2 & 12,000 & $\LACP$ & 80,000k & $\infty$ & 1 \hse& 579 & 42 & 38,444k\\
$\RS'_{13}$ & $\PSP$ & 29 & 452 & 2 & 100 & $\LACT$ & $\infty$ & 100k & 4 \hse& 259 & 5 & 2,957k\\\midrule
\multicolumn{2}{l}{$\RS'_{1-6}$} & 77 & 603 & 4\\
\multicolumn{2}{l}{$\RS'_{7-13}$} & 82 & 596 & 3\\
\multicolumn{2}{l}{$\RS'_{1-13}$} & 94 & 627 & 4\\
\multicolumn{2}{l}{$\Rbase,\RS'_{1-13}$} & 94 & 648 & 5\\ 
\end{tabular}
\end{table}

\FloatBarrier
\clearpage
\section{Supplementary Material for Section~\ref{sec-dag-synthesis}: Precision
Plots}
\label{app-supp-dag-synthesis-plots}

Figure~\ref{fig-precision-full} below extends Fig.~\ref{fig-precision-main} in
Sect.~\ref{sec-dag-synthesis} by plots based on further sequences $\DLEM$ of
proof terms specified in Table~\ref{tab-lemma-settings}.

\medskip

\newcommand{\xsubcaptionbox}[2]
{\begin{minipage}{0.3\textwidth}
    \centering\small #2\\
    #1
\end{minipage}}

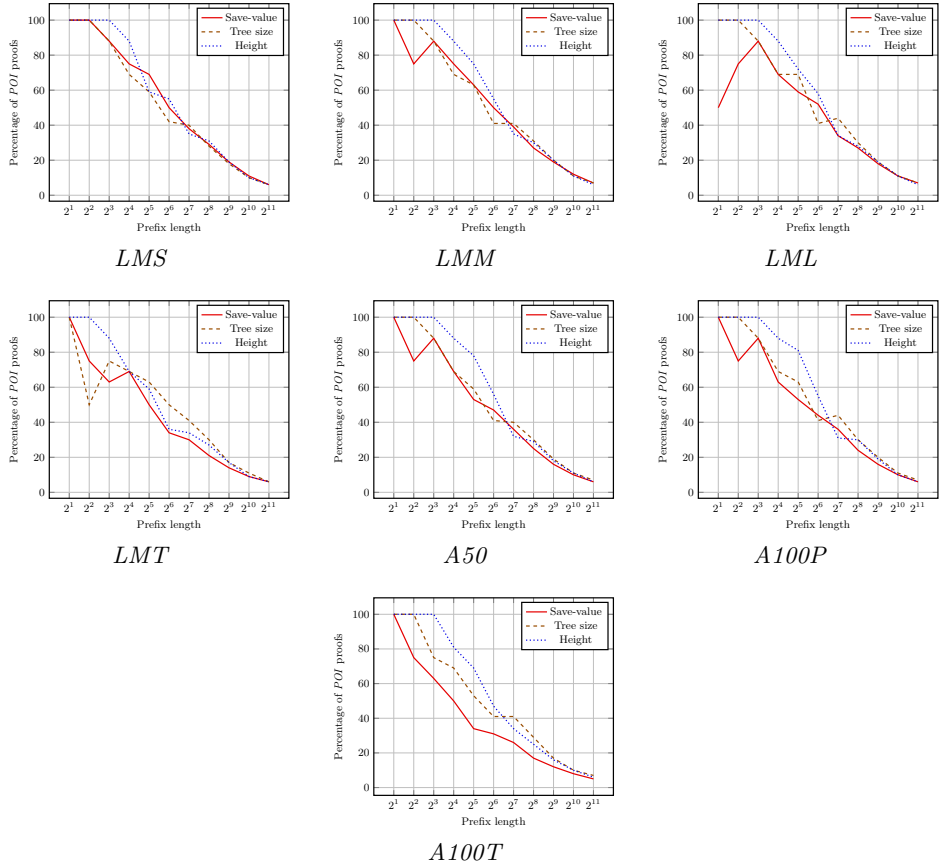
\begin{figure}[h]
  \centering\small

\xsubcaptionbox{$\LMS$}{ %
\scalebox{0.46}{
\begin{tikzpicture}
  \begin{semilogxaxis}[
      cycle list name=prec,
      log basis x = {2},
      xmajorgrids = true,
      ymajorgrids = true,
      line width=1.0,
      xtick = {2,4,8,16,32,64,128,256,512,1024,2048},
      xlabel={Prefix length}, ylabel={Percentage of \POI proofs},
    ]
    \addplot coordinates {(2,100)(4,100)(8,88)(16,75)(32,69)(64,50)(128,38)(256,29)(512,19)(1024,11)(2048,6)};
    \addplot coordinates {(2,100)(4,100)(8,88)(16,69)(32,59)(64,42)(128,40)(256,28)(512,18)(1024,10)(2048,6)};
    \addplot coordinates {(2,100)(4,100)(8,100)(16,88)(32,59)(64,55)(128,35)(256,31)(512,19)(1024,10)(2048,6)};
    \legend{{Save-value},{Tree size},{Height}}
\end{semilogxaxis}
\end{tikzpicture}}}
\hfill
\xsubcaptionbox{$\LMM$}{ %
\scalebox{0.46}{
\begin{tikzpicture}
  \begin{semilogxaxis}[
      cycle list name=prec,      
      log basis x = {2},
      xmajorgrids = true,
      ymajorgrids = true,
      line width=1,
      xtick = {2,4,8,16,32,64,128,256,512,1024,2048},
      xlabel={Prefix length}, ylabel={Percentage of \POI proofs},      
    ]
    \addplot coordinates {(2,100)(4,75)(8,88)(16,75)(32,63)(64,50)(128,39)(256,27)(512,19)(1024,12)(2048,7)};
    \addplot coordinates {(2,100)(4,100)(8,88)(16,69)(32,63)(64,41)(128,41)(256,31)(512,20)(1024,11)(2048,7)};
    \addplot coordinates {(2,100)(4,100)(8,100)(16,88)(32,75)(64,55)(128,35)(256,30)(512,20)(1024,11)(2048,6)};
    \legend{{Save-value},{Tree size},{Height}}
\end{semilogxaxis}
\end{tikzpicture}}}
\hfill
\xsubcaptionbox{$\LML$}{  %
\scalebox{0.46}{
\begin{tikzpicture}
  \begin{semilogxaxis}[
      cycle list name=prec,      
      log basis x = {2},
      xmajorgrids = true,
      ymajorgrids = true,
      line width=1,
      xtick = {2,4,8,16,32,64,128,256,512,1024,2048},
      xlabel={Prefix length}, ylabel={Percentage of \POI proofs},
    ]
    \addplot coordinates {(2,50)(4,75)(8,88)(16,69)(32,59)(64,52)(128,34)(256,27)(512,18)(1024,11)(2048,7)};
    \addplot coordinates {(2,100)(4,100)(8,88)(16,69)(32,69)(64,41)(128,44)(256,30)(512,19)(1024,11)(2048,7)};
    \addplot coordinates {(2,100)(4,100)(8,100)(16,88)(32,72)(64,58)(128,34)(256,28)(512,19)(1024,11)(2048,6)};
    \legend{{Save-value},{Tree size},{Height}}
\end{semilogxaxis}
\end{tikzpicture}}}

\bigskip

\xsubcaptionbox{$\LMT$}{ %
\scalebox{0.46}{
\begin{tikzpicture}
  \begin{semilogxaxis}[
      cycle list name=prec,
      log basis x = {2},
      xmajorgrids = true,
      ymajorgrids = true,
      line width=1,
      xtick = {2,4,8,16,32,64,128,256,512,1024,2048},
      xlabel={Prefix length}, ylabel={Percentage of \POI proofs},
    ]
    \addplot coordinates {(2,100)(4,75)(8,63)(16,69)(32,50)(64,34)(128,30)(256,21)(512,14)(1024,9)(2048,6)};
    \addplot coordinates {(2,100)(4,50)(8,75)(16,69)(32,63)(64,50)(128,41)(256,30)(512,17)(1024,11)(2048,6)};
    \addplot coordinates {(2,100)(4,100)(8,88)(16,69)(32,59)(64,36)(128,34)(256,27)(512,17)(1024,9)(2048,6)};
    \legend{{Save-value},{Tree size},{Height}}
\end{semilogxaxis}
\end{tikzpicture}}}
\hfill
\xsubcaptionbox{$\LAL$}{  %
\scalebox{0.46}{
\begin{tikzpicture}
  \begin{semilogxaxis}[
      cycle list name=prec,
      log basis x = {2},
      xmajorgrids = true,
      ymajorgrids = true,
      line width=1,
      xtick = {2,4,8,16,32,64,128,256,512,1024,2048},
      xlabel={Prefix length}, ylabel={Percentage of \POI proofs},
    ]
    \addplot coordinates {(2,100)(4,75)(8,88)(16,69)(32,53)(64,47)(128,36)(256,25)(512,16)(1024,10)(2048,6)};
    \addplot coordinates {(2,100)(4,100)(8,88)(16,69)(32,59)(64,41)(128,40)(256,30)(512,19)(1024,11)(2048,7)};
    \addplot coordinates {(2,100)(4,100)(8,100)(16,88)(32,78)(64,56)(128,32)(256,29)(512,18)(1024,11)(2048,6)};
    \legend{{Save-value},{Tree size},{Height}}
\end{semilogxaxis}
\end{tikzpicture}}}
\hfill
\xsubcaptionbox{$\LACP$}{  %
\scalebox{0.46}{
\begin{tikzpicture}
  \begin{semilogxaxis}[
      cycle list name=prec,
      log basis x = {2},
      xmajorgrids = true,
      ymajorgrids = true,
      line width=1,
      xtick = {2,4,8,16,32,64,128,256,512,1024,2048},
      xlabel={Prefix length}, ylabel={Percentage of \POI proofs},
    ]
    \addplot coordinates {(2,100)(4,75)(8,88)(16,63)(32,53)(64,44)(128,36)(256,24)(512,16)(1024,10)(2048,6)};
    \addplot coordinates {(2,100)(4,100)(8,88)(16,69)(32,63)(64,41)(128,44)(256,30)(512,20)(1024,11)(2048,7)};
    \addplot coordinates {(2,100)(4,100)(8,100)(16,88)(32,81)(64,55)(128,31)(256,30)(512,19)(1024,10)(2048,6)};
    \legend{{Save-value},{Tree size},{Height}}
\end{semilogxaxis}
\end{tikzpicture}}}

\bigskip

\xsubcaptionbox{$\LACT$}{ %
\scalebox{0.46}{
\begin{tikzpicture}
  \begin{semilogxaxis}[
      cycle list name=prec,
      log basis x = {2},
      xmajorgrids = true,
      ymajorgrids = true,
      line width=1,
      xtick = {2,4,8,16,32,64,128,256,512,1024,2048},
      xlabel={Prefix length}, ylabel={Percentage of \POI proofs},
    ]
    \addplot coordinates {(2,100)(4,75)(8,63)(16,50)(32,34)(64,31)(128,26)(256,17)(512,12)(1024,8)(2048,5)};
    \addplot coordinates {(2,100)(4,100)(8,75)(16,69)(32,53)(64,41)(128,41)(256,29)(512,17)(1024,10)(2048,7)};
    \addplot coordinates {(2,100)(4,100)(8,100)(16,81)(32,69)(64,47)(128,34)(256,25)(512,16)(1024,10)(2048,6)};
    \legend{{Save-value},{Tree size},{Height}}
\end{semilogxaxis}
\end{tikzpicture}}}

\caption{Percentage of prefix members that prove a \POI theorem for sequences
  of proof terms obtained by lemma synthesis as specified in
  Table~\ref{tab-lemma-settings}, ordered by different criteria.}

\label{fig-precision-full}

\end{figure}

\FloatBarrier
\clearpage
\section{Supplementary Material for Section~\ref{sec-dag-synthesis}: Bar Chart
  on \Vampire with Lemmas}

\begin{figure}[h]
\scalebox{0.8}{
\begin{tikzpicture}
  \begin{axis}[
      xbar stacked,
      axis x line*=bottom,
      axis y line*=none,
      y axis line style = { draw = none },
    height = 3.5cm,
    width = 14.5cm,
    bar width=0.4cm,
    xmin=-1,
    xmax=1374,
    xtick distance = 200,
    enlarge y limits  = 0.4,
    nodes near coords,
    extra x ticks = {1374},
    symbolic y coords={{60 s},{600 s},{3,600 s}},
    nodes near coords,    
    point meta=x,    
  ]
    \addplot [fill=blue!10] coordinates {(538,{60 s})(939,{600 s})(1023,{3,600 s})};
    \addplot [fill=red!25] coordinates {(292,{60 s})(291,{600 s})(267,{3,600 s})};
  \end{axis}
\end{tikzpicture}}
\caption{Increase of \POI problems solved by \Vampire in 3,600~s, 600~s and
  60~s through supplying the lemma set from $\RS_3$.}
\label{fig-vampire-lemmas}
\end{figure}
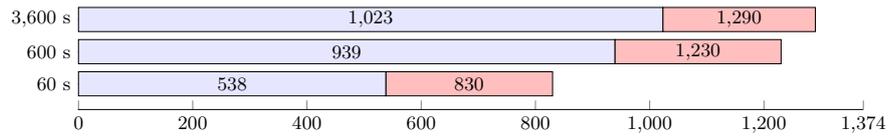

Figure~\ref{fig-vampire-lemmas} below supplements
Table~\ref{tab-lemmas-for-provers} in Sect.~\ref{sec-dag-synthesis}
Fig~\ref{fig-vampire-lemmas} with information about the solving times for the
best overall result, \Vampire with the lemmas from $\RS_{3}$. It shows that
the performance gain through the lemmas does not only occur with the long
timeouts of 3,600~s.

\medskip

\FloatBarrier
\clearpage
\section{Supplementary Material for Section~\ref{sec-combinatory}: Finding
  Combinatory Schemas in \SETMM}
\label{app-finding-combs}

When invoked with level characterization $\PP$ or $\PK$, \SGCD takes a
configured set of proof schemas, i.e., proof patterns and direct proof
schemas, each characterized by a combinator and an arity. The question arises,
which combinators might be beneficial here. To get some orientation, we looked
at the proofs of \SETMM. Each proof of a theorem may be naturally viewed as a
production $t(V_1,\ldots,V_n) \pro d[V_1,\ldots,V_n]$ in a tree grammar
\cite{cw:zz:2025} (see also \appref{app-implication-form}), where the
nonterminal is the theorem name and $V_1,\ldots,V_n$ are parameters (i.e.,
proof term variables) that occur in $d$. Now, we eliminate the variables from
$d$ with a $\lambda$ to SKI translation of $\lambda V_1\ldots V_n . d$, and
record as combinators the occurrences of maximal subterms that only have
primitive combinators as constants. The (speculative) intuition is that these
provide an inventory of the ways in which human experts structure proofs.

For the $\lambda$ to SKI translation we followed \cite{peytonjones:87}, and
hence considered as primitive combinators those resulting from this method,
shown in Table~\ref{tab-combinators-full}.

\begin{table}[h]
  \centering
\caption{Considered primitive combinators with defining $\lambda$-term and
  principal type. $\c{S_4}$, $\c{B_4}$, $\c{C_4}$ are called $\c{S'}$,
  $\c{B^*}$,$\c{C'}$ in \cite{peytonjones:87}.}
\label{tab-combinators-full}
$\begin{array}{lll}
        & \text{$\lambda$-Term} & \text{Principal Type} \\\midrule
  \c{S} & \lambda  V_1V_2V_2  \ldot \D(\D(V_1, V_3), \D(V_2, V_3))
  & (x\oimp (y\oimp z))\oimp ((x\oimp y)\oimp (x\oimp z))\\
  \c{K} & \lambda V_1V_2 \ldot V_1
  & x\oimp (y\oimp x)\\
  \c{I} & \lambda V_1 \ldot V_1
  & x\oimp x\\  
  \c{B} & \lambda V_1V_2V_3 \ldot \D(V_1, \D(V_2, V_3)) 
  & (x\oimp y)\oimp ((z\oimp x)\oimp (z\oimp y))\\
  \c{C} & \lambda V_1V_2V_2 \ldot \D(\D(V_1, V_3), V_2) 
  & (x\oimp (y\oimp z))\oimp (y\oimp (x\oimp z))\\
  \c{S_4} &  \lambda V_1V_2V_3V_4\; .   & (x\oimp (y\oimp z))\oimp\\
          & \D(\D(V_1, \D(V_2, V_4)), \D(V_3, V_4)) 
  & \hspace{5em} ((u\oimp x)\oimp ((u\oimp y)\oimp (u\oimp z)))\\
  \c{B_4} &  \lambda V_1V_2V_3V_4 . \D(V_1, \D(V_2, \D(V_3, V_4)))
  & (x\oimp y)\oimp ((z\oimp x)\oimp ((u\oimp z)\oimp (u\oimp y)))\\
  \c{C_4} & \lambda V_1V_2V_3V_4 . \D(\D(V_1, \D(V_2, V_4)), V_3)
  & (x\oimp (y\oimp z))\oimp ((u\oimp x)\oimp (y\oimp (u\oimp z)))\\  
\end{array}$
\end{table}

We then counted combinator occurrences in the proofs of the first two thirds
of \SETMM with 28,673 theorems, the best curated part that was also used to
extract the \POI theorems. We found 3,424 combinators with more than a single
occurrence, and 187 ones with more than 100 occurrences. The 21 combinators
with the most occurrences are shown in Table~\ref{tab-comb-freq}. At positions
9,10,11,13,15,18,21 we find instances of the bulk combinators used in the
linear $\lambda$ to SKI translation \cite{kiselyov:lambdaToSKI:2018}.
Candidate sets of combinators for use as proof schemas were finally determined
by comparing for up to four combinators among the 19 most frequent ones the
number of theorems in whose proof (i.e., the converted production) all of them
occur.

\clearpage
\ 
\begin{table}[t]
  \centering
  \caption{The 21 combinators that occur most frequently in the proofs of the
    first two thirds of \SETMM, when viewed as combinator terms. The right
    column shows the number of occurrences.}
  \smallskip
  \label{tab-comb-freq}
\begin{tabular}{B@{\hspace{4pt}}A@{\hspace{0.5em}}B}
1. & \c{B_{4}} & 169,394 \\
2. & \c{C} & 164,919 \\
3. & \c{C_{4}} & 134,575 \\
4. & \c{B} & 91,964 \\
5. & \c{B_{4}} \c{C} & 81,257 \\
6. & \c{S_{4}} & 75,918 \\
7. & \c{B_{4}} \c{S} & 44,722 \\
\end{tabular}
\hspace{\fill}
\begin{tabular}{B@{\hspace{4pt}}A@{\hspace{0.5em}}B}
8. & \c{S} & 39,339 \\
9. & \c{C_{4}} \c{C} & 30,110 \\
10. & \c{C_{4}} (\c{C_{4}} \c{C}) & 13,044\\
11. & \c{C_{4}} (\c{C_{4}} (\c{C_{4}} \c{C})) & 10,458 \\
12. & \c{B_{4}} \c{B_{4}} & 6,911 \\
13. & \c{S_{4}} \c{S} & 6,602 \\
14. & \c{C_{4}} \c{B_{4}} & 6,251 \\
\end{tabular}
\hspace{\fill}
\begin{tabular}{B@{\hspace{4pt}}A@{\hspace{0.5em}}B}
15. & \c{C_{4}} (\c{C_{4}} (\c{C_{4}} (\c{C_{4}} \c{C}))) & 5,541 \\
16. & \c{B_{4}} (\c{C_{4}} \c{C}) & 5,318 \\
17. & \c{B_{4}} \c{C_{4}} & 4,216 \\
18. & \c{C_{4}} (\c{C_{4}} (\c{C_{4}} (\c{C_{4}} (\c{C_{4}} \c{C})))) & 3,575 \\
19. & \c{B_{4}} \c{B} & 3,062 \\
20. & \c{I} & 3,017 \\
21. & \c{S_{4}} (\c{S_{4}} \c{S}) & 2,594
\end{tabular}

\end{table}

\FloatBarrier
\clearpage
\section{Generated Theorems That Seem Hard for Provers}
\label{app-hard-problems}

Table~\ref{tab-hard-problems} below lists those 12 \POI theorems with rating 1
(as specified in Sect.~\ref{sec-ideal}), for which proofs were in the results
of runs $\RBP$, $\RBT$ (Table~\ref{tab-base-pt}), $\RS_1{-}\RS_{13}$
(Table~\ref{tab-results-sav}) or $\RC_1{-}\RC_{22}$
(Table~\ref{tab-results-pp} and~\ref{tab-results-pk}). The corresponding proof
problems, proving the theorem statement from the three axioms, were given
again to the provers, now on a Linux server platform with four Intel Xeon
E5-4640 processors and 256 GB RAM, and with a larger time limit of 15,000~s.
Some problems were provable in a shorter time than the 3,600~s from the
cluster runs, possibly due to a different strategy scheduling by the provers
that were given the time limits as arguments, and to the availability of
concurrent processing units.

\begin{table}
  \caption{\POI theorems of rating 1 for which proofs were encountered in
    theorem-generating experiments \h{Runs}. The prover-labeled columns show
    the solving time of the prover in seconds. A hyphen indicates timeout.
    Versions and settings of the provers were as in Sect.~\ref{sec-ideal},
    except for \EProverASC, which refers to \EProver with option
    \code{--auto-schedule=10}.}
  \label{tab-hard-problems}
  \centering \small
  \setlength{\tabcolsep}{4pt}

  \smallskip
  
  \begin{tabular}{llrrr}
  \h{\POI theorem} & \h{Runs} & \Vampire & \EProver & \EProverASC\\\midrule
  \poi{3anim1i} & $\RBP,\RS_3,\RS_6,\RS_{13},\RC_1,\RC_3,\RC_4$   & 1,107    & 5,166 & 9,996\\
  \poi{exp41}   & $\RC_2,\RC_3,\RC_4,\RC_5,\RC_6,\RC_{11},\RC_{13}$  & --       & --    & --\\    
  \poi{exp4b}   & $\RBP$                            & 11,300   & 8,648 & 4,012\\
  \poi{imp41}   & $\RC_1,\RC_2,\RC_3,\RC_4$                & 136      & 14,120 & --\\
  \poi{mp2and}  & $\RS_{12}$                           & 555      & --     & 13,727\\   
  \poi{mp3anl3} & $\RS_6,\RC_2,\RC_5$                    & --       & 11,422 & 11,269\\
  \poi{mpanl12} & $\RS_{10},\RC_2,\RC_5,\RC_6,\RC_{20},\RC_{21}$ & 141      & --     & --\\
  \poi{mpanl2}  & $\RS_6,\RC_2$                        & 11,323   & 9,708    & 7,702\\
  \poi{stoic3}  & $\RBP,\RBT,\RS_{1-13},\RC_1,\RC_3,\RC_4,\RC_6$ & 11,323 & 4,478 & 4,997\\
  \poi{syland}  & $\RS_{13}$                           & 11,733   & --        & --\\
  \poi{sylani}  & $\RS_3,\RS_6,\RS_8,\RS_{12},\RS_{13},\RC_6$      & 11,299   & 9,876     & 7,773\\
  \poi{sylanl1} & $\RBP$ & -- & -- & --
\end{tabular}
\end{table}

In the following subsections we show for each of the \POI theorems from
Table~\ref{tab-hard-problems} one of the proofs obtained in our experiments.
We write the proof terms as DAG grammars.

\subsection{Proofs by Run $\RBP$}

\newenvironment{showproof}{\setlength{\parindent}{0pt}\small\rule{\textwidth}{1pt}\medskip}{\medskip}

\begin{showproof}
\begin{tabular}{l@{\hspace{1em}}l}
\POI Theorem   & \poi{3anim1i}\\
Run            & $\RBP$\\
Compacted size & 29\\
Tree size      & 637\\
Height         & 29\\
MGT            & $(x\oimp y)\oimp(\fn((\fn((x\oimp z))\oimp u))\oimp\fn((\fn((y\oimp z))\oimp u)))$
\end{tabular}

\medskip

$\begin{array}{rcl}
4 & \pro & \D(2, \D(1, 2))\\
5 & \pro & \D(4, 1)\\
6 & \pro & \D(5, \D(2, 5))\\
7 & \pro & \D(6, 1)\\
8 & \pro & \D(7, 1)\\
9 & \pro & \D(\D(7, 8), 7)\\
10 & \pro & \D(\D(7, \D(4, \D(9, 3))), 8)\\
11 & \pro & \D(\D(5, \D(9, \D(2, \D(\D(7, \D(\D(6, 10), 3)), 3)))), 7)\\
12 & \pro & \D(11, \D(\D(10, \D(4, \D(\D(5, 11), 5))), 5))\\
13 & \pro & \D(\D(5, 12), 12)\\
\end{array}$
\end{showproof}

\begin{showproof}
\begin{tabular}{l@{\hspace{1em}}l}
  \POI theorem & \poi{exp4b}\\
  Run            & $\RBP$\\  
  Compacted size & 25\\
  Tree size & 487\\
  Height & 25\\
  MGT &  $(\fn((x\oimp\fn(y)))\oimp(\fn((z\oimp\fn(u)))\oimp v))\; \oimp$\\
  & $(x\oimp(y\oimp(z\oimp(u\oimp v))))$
\end{tabular}  

\medskip

$\begin{array}{rcl}
4 & \pro & \D(2, \D(1, 2))\\
5 & \pro & \D(4, 1)\\
6 & \pro & \D(5, \D(2, 5))\\
7 & \pro & \D(6, 1)\\
8 & \pro & \D(7, 1)\\
9 & \pro & \D(\D(7, 8), 7)\\
10 & \pro & \D(5, \D(9, \D(2, \D(\D(7, \D(\D(6, \D(\D(7, \D(4, \D(9, 3))), 8)), 3)), 3))))\\
11 & \pro & \D(5, \D(\D(7, \D(10, 5)), 10))\\
12 & \pro & \D(11, 11)\\
\end{array}$
\end{showproof}

\begin{showproof}
\begin{tabular}{l@{\hspace{1em}}l}
  \POI theorem & \poi{stoic3}\\
  Run            & $\RBP$\\    
  Compacted size &  24\\
  Tree size &  183\\
  Height &  24\\
  MGT & $(x\oimp y)\oimp((\fn((y\oimp z))\oimp u)\oimp(\fn((x\oimp z))\oimp u))$
\end{tabular}

\medskip

$\begin{array}{rcl}
4 & \pro & \D(2, \D(1, 2))\\
5 & \pro & \D(4, 1)\\
6 & \pro & \D(5, \D(2, 5))\\
7 & \pro & \D(6, 1)\\
8 & \pro & \D(7, 1)\\
9 & \pro & \D(6, \D(\D(7, \D(4, \D(\D(\D(7, 8), 7), 3))), 8))\\
10 & \pro & \D(\D(5, \D(\D(7, \D(\D(5, \D(5, \D(\D(7, \D(9, 3)), 3))), 7)), 9)), 7)\\
\end{array}$
\end{showproof}

\begin{showproof}
\begin{tabular}{l@{\hspace{1em}}l}
  \POI theorem & \poi{sylanl1}\\
  Run            & $\RBP$\\    
  Compacted size &  25\\
  Tree size &  357\\
  Height &  25\\
  MGT &  $(x\oimp y)\oimp$\\
  & $((\fn((\fn((y\oimp z))\oimp u))\oimp v)\oimp(\fn((\fn((x\oimp z))\oimp u))\oimp v))$
\end{tabular}

\medskip

$\begin{array}{rcl}
4 & \pro & \D(2, \D(1, 2))\\
5 & \pro & \D(4, 1)\\
6 & \pro & \D(5, \D(2, 5))\\
7 & \pro & \D(6, 1)\\
8 & \pro & \D(7, 1)\\
9 & \pro & \D(6, \D(\D(7, \D(4, \D(\D(\D(7, 8), 7), 3))), 8))\\
10 & \pro & \D(5, \D(\D(7, \D(\D(5, \D(5, \D(\D(7, \D(9, 3)), 3))), 7)), 9))\\
11 & \pro & \D(10, \D(10, 7))\\
\end{array}$
\end{showproof}

\subsection{Proofs by Runs $\RS_{1-13}$}

References to lemmas in the obtained proofs are replaced here with the proofs
of the lemmas, to obtain full proofs of the theorems just from the three
original axioms.

\begin{showproof}
\begin{tabular}{l@{\hspace{1em}}l}
  \POI theorem & \poi{mp2and}\\
  Run & $\RS_{12}$\\
  Compacted size & 30\\
  Tree size & 668\\
  Height & 25\\
  MGT & $(x\oimp y)\oimp((x\oimp z)\oimp((x\oimp(\fn((y\oimp\fn(z)))\oimp u))\oimp(x\oimp u)))$
\end{tabular}

\medskip

$\begin{array}{rcl}
4 & \pro & \D(2, \D(1, 2))\\
5 & \pro & \D(4, 1)\\
6 & \pro & \D(\D(5, \D(2, 5)), 1)\\
7 & \pro & \D(6, 1)\\
8 & \pro & \D(\D(5, \D(\D(6, 7), 6)), 2)\\
9 & \pro & \D(8, 8)\\
10 & \pro & \D(9, \D(\D(9, 8), 3))\\
11 & \pro & \D(\D(\D(5, \D(5, \D(\D(5, 4), 2))), 5),\D(10, \D(\D(\D(8, \D(7, \D(10, \D(6, \D(2,\\&& \D(10, 7)))))), 8), 1)))\\
\end{array}$
\end{showproof}

\begin{showproof}
\begin{tabular}{l@{\hspace{1em}}l}
  \POI theorem & \poi{mp3anl3}, \poi{mpanl2} (the MGT subsumes both theorem formulas)\\
  Run & $\RS_{6}$\\  
  Compacted size & 42\\
  Tree size & 95\\
  Height & 17\\
  MGT & $x\oimp((\fn((\fn((y\oimp\fn(x)))\oimp z))\oimp u)\oimp(\fn((y\oimp z))\oimp u))$\\
\end{tabular}

\medskip

$\begin{array}{rcl}
4 & \pro & \D(2, \D(1, 3))\\
5 & \pro & \D(2, \D(\D(2, \D(1, 2)), 1))\\
6 & \pro & \D(2, \D(4, \D(4, 1)))\\
7 & \pro & \D(2, \D(1, \D(\D(2, \D(1, 5)), 1)))\\
8 & \pro & \D(\D(2, \D(1, \D(\D(2, 2), 6))), 1)\\
9 & \pro & \D(4, 8)\\
10 & \pro & \D(7, \D(\D(2, \D(1, \D(\D(2, \D(1, \D(4, \D(2, \D(1, \D(3, \D(6, 1))))))), 8))),\D(7,\\&& \D(\D(2, \D(1, 9)), \D(\D(\D(2, \D(1, \D(5, \D(1, 1)))), 2), 9)))))\\
\end{array}$
\end{showproof}

\begin{showproof}
\begin{tabular}{l@{\hspace{1em}}l}
  \POI theorem & \poi{mpanl12}\\
  Run & $\RS_{10}$\\  
  Compacted size & 26\\
  Tree size & 63\\
  Height & 13\\
  MGT & $x\oimp(y\oimp((\fn((\fn((x\oimp\fn(y)))\oimp\fn(z)))\oimp
  u)\oimp(z\oimp u)))$
\end{tabular}

\medskip

$\begin{array}{rcl}
4 & = & \D(2, \D(1, 3))\\
5 & = & \D(\D(2, \D(1, 2)), 1)\\
6 & = & \D(4, \D(\D(2, \D(1, \D(2, \D(\D(2, \D(4, \D(4, 1))), 1)))), 1))\\
7 & = & \D(\D(\D(\D(5, \D(2, \D(\D(5, 5), 5))), 1), 6), \D(\D(5, \D(\D(5, \D(2, 5)), 1)), 6))\\
\end{array}$
\end{showproof}

\begin{showproof}
\begin{tabular}{l@{\hspace{1em}}l}
  \POI theorem & \poi{syland}\\
  Run & $\RS_{13}$\\  
  Compacted size & 35\\
  Tree size & 58\\
  Height & 18\\
  MGT & $(x\oimp(y\oimp z))\; \oimp$\\
  & $((x\oimp(\fn((z\oimp u))\oimp v))\oimp(x\oimp(\fn((y\oimp u))\oimp v)))$
\end{tabular}

\medskip

$\begin{array}{rcl}
4 & \pro & \D(2, \D(1, 2))\\
5 & \pro & \D(2, \D(1, 3))\\
6 & \pro & \D(4, 1)\\
7 & \pro & \D(2, \D(5, \D(5, 1)))\\
8 & \pro & \D(\D(2, \D(1, \D(2, 6))), 1)\\
9 & \pro & \D(\D(\D(2, \D(1, 4)), 6), \D(\D(2, \D(1, 8)), \D(\D(2, \D(1, \D(\D(2, \D(1, \D(5,\\
         && \D(2, \D(1, \D(3, \D(7, 1))))))), \D(\D(2, \D(1, \D(\D(2, 2), 7))), 1)))), 8)))\\
\end{array}$
\end{showproof}

\begin{showproof}
\begin{tabular}{l@{\hspace{1em}}l}
  \POI theorem & \poi{sylani}\\
  Run & $\RS_{13}$\\    
  Compacted size & 34\\
  Tree size & 55\\
  Height & 18\\
  MGT & $(x\oimp y)\oimp((z\oimp(\fn((y\oimp u))\oimp v))\oimp(z\oimp(\fn((x\oimp u))\oimp v)))$
\end{tabular}

\medskip

$\begin{array}{rcl}
4 & \pro & \D(2, \D(1, 3))\\
5 & \pro & \D(\D(2, \D(1, 2)), 1)\\
6 & \pro & \D(2, \D(4, \D(4, 1)))\\
7 & \pro & \D(\D(2, \D(1, \D(2, 5))), 1)\\
8 & \pro & \D(\D(2, \D(1, 5)), \D(\D(2, \D(1, 7)), \D(\D(2, \D(1, \D(\D(2, \D(1, \D(4, \D(2, \D(1,\\
&& \D(3, \D(6, 1))))))), \D(\D(2, \D(1, \D(\D(2, 2), 6))), 1)))), 7)))\\
\end{array}$
\end{showproof}

\subsection{Proofs by Run $\RC_{2}$}
\label{appsec-pattern-proofs}

This run used for proof term construction the $\D$ function (expressed via
$\c{I}/2$) along with the following patterns, shown here as rewrite rules to
combinator terms that specify their semantics:

\[
\begin{array}{lcl}
  \fp_1(V_1, V_2, V_3) & \defpatt & \D(\D(\D(\c{B_{4}}, V_1), V_2), V_3)\\
  \fp_2(V_1, V_2)    & \defpatt & \D(\D(\c{C}, V_1), V_2)\\
  \fp_3(V_1, V_2)    & \defpatt & \D(\D(\c{S}, V_1), V_2)
\end{array}
\]

We show the proofs in three forms, the latter two obtained in succession in
linear time with linear size increase: (1) the pattern proof term as DAG
grammar returned by \SGCD; (2) the proof term obtained by rewriting the
patterns with the above rules into combinator terms; (3) the proof term if the
primitive combinators are replaced by proofs of their MGTs that only refer to
the three original axioms.

\begin{showproof}
\begin{tabular}{l@{\hspace{1em}}l}
  \POI theorem & \poi{exp41}\\
  Run & $\RC_{2}$\\    
  MGT & $(\fn((\fn((\fn((x\oimp\fn(y)))\oimp\fn(z)))\oimp\fn(u)))\oimp v)\oimp$\\&$(x\oimp(y\oimp(z\oimp(u\oimp v))))$
\end{tabular}

\subsubsection{(1) Pattern Proof}\ \smallskip

\begin{tabular}{l@{\hspace{1em}}l}
  Compacted size & 18\\
  Tree size & 199\\
  Height & 18\\
\end{tabular}

\medskip

$\begin{array}{rcl}
4 & \pro & \fp_1(3, 3, 1)\\
5 & \pro & \fp_2(\fp_1(1, 3, \D(2, \fp_1(3, \D(2, 4), 1))), 1)\\
6 & \pro & \D(2, \D(1, 5))\\
7 & \pro & \fp_3(\D(1, \fp_2(\fp_1(6, 2, 1), 4)), 2)\\
8 & \pro & \fp_1(3, 7, \fp_1(3, \D(6, \fp_1(\fp_1(3, 7, 6), 3, 7)), 5))\\
9 & \pro & \fp_1(8, 8, 8)
\end{array}$

\subsubsection{(2) Proof with Patterns Rewritten to Combinators}\ \smallskip

\begin{tabular}{l@{\hspace{1em}}l}
   Compacted size & 34\\
   Tree size & 405\\
   Height & 32\\
\end{tabular}

\medskip

$\begin{array}{rcl}
4 & \pro & \D(\c{B_{4}}, 3)\\
5 & \pro & \D(\D(4, 3), 1)\\
6 & \pro & \D(\D(\c{C}, \D(\D(\D(\c{B_{4}}, 1), 3), \D(2, \D(\D(4, \D(2, 5)), 1)))), 1)\\
7 & \pro & \D(2, \D(1, 6))\\
8 & \pro & \D(\D(\c{S}, \D(1, \D(\D(\c{C}, \D(\D(\D(\c{B_{4}}, 7), 2), 1)), 5))), 2)\\
9 & \pro & \D(4, 8)\\
10 & \pro & \D(9, \D(\D(4, \D(7, \D(\D(\D(\c{B_{4}}, \D(9, 7)), 3), 8))), 6))\\
11 & \pro & \D(\D(\D(\c{B_{4}}, 10), 10), 10)\\
\end{array}$

\subsubsection{(3) Final Converted Proof from the Original Axioms}\ \smallskip

\begin{tabular}{l@{\hspace{1em}}l}
    Compacted size & 50\\
    Tree size & 1,583\\
    Height & 40\\
\end{tabular}

\medskip

$\begin{array}{rcl}
4 & \pro & \D(1, 1)\\
5 & \pro & \D(\D(2, \D(1, 2)), 1)\\
6 & \pro & \D(\D(2, \D(1, \D(\D(2, 2), 4))), \D(\D(2, 4), 2))\\
7 & \pro & \D(\D(2, \D(1, \D(2, \D(1, 5)))), 5)\\
8 & \pro & \D(7, 3)\\
9 & \pro & \D(\D(8, 3), 1)\\
10 & \pro & \D(\D(6, \D(\D(\D(7, 1), 3), \D(2, \D(\D(8, \D(2, 9)), 1)))), 1)\\
11 & \pro & \D(2, \D(1, 10))\\
12 & \pro & \D(\D(2, \D(1, \D(\D(6, \D(\D(\D(7, 11), 2), 1)), 9))), 2)\\
13 & \pro & \D(8, 12)\\
14 & \pro & \D(13, \D(\D(8, \D(11, \D(\D(\D(7, \D(13, 11)), 3), 12))), 10))\\
15 & \pro & \D(\D(\D(7, 14), 14), 14)\\
\end{array}$

\end{showproof}

\begin{showproof}
\begin{tabular}{l@{\hspace{1em}}l}
  \POI theorem & \poi{imp41}\\
  Run & $\RC_{2}$\\    
  MGT & $(x\oimp(y\oimp(z\oimp(u\oimp v))))\;\oimp$\\
      &$(\fn((\fn((\fn((x\oimp\fn(y)))\oimp\fn(z)))\oimp\fn(u)))\oimp v)$
\end{tabular}

\subsubsection{(1) Pattern Proof}\ \smallskip

\begin{tabular}{l@{\hspace{1em}}l}
  Compacted size & 17\\
  Tree size & 196\\
  Height & 17\\
\end{tabular}

\medskip

$\begin{array}{rcl}
4 & \pro & \fp_1(3, 3, 1)\\
5 & \pro & \fp_2(\fp_1(1, 3, \D(2, \fp_1(3, \D(2, 4), 1))), 1)\\
6 & \pro & \D(2, \D(1, 5))\\
7 & \pro & \fp_3(\D(1, \fp_2(\fp_1(6, 2, 1), 4)), 2)\\
8 & \pro & \fp_1(\D(6, \fp_1(\fp_1(3, 7, 6), 3, 7)), 5, 7)\\
9 & \pro & \fp_1(8, 8, 8)
\end{array}$

\subsubsection{(2) Proof with Patterns Rewritten to Combinators}\ \smallskip

\begin{tabular}{l@{\hspace{1em}}l}
  Compacted size & 34\\
  Tree size & 396\\
  Height & 32
\end{tabular}

\medskip

$\begin{array}{rcl}
4 & \pro & \D(\c{B_{4}}, 3)\\
5 & \pro & \D(\D(4, 3), 1)\\
6 & \pro & \D(\D(\c{C}, \D(\D(\D(\c{B_{4}}, 1), 3), \D(2, \D(\D(4, \D(2, 5)), 1)))), 1)\\
7 & \pro & \D(2, \D(1, 6))\\
8 & \pro & \D(\D(\c{S}, \D(1, \D(\D(\c{C}, \D(\D(\D(\c{B_{4}}, 7), 2), 1)), 5))), 2)\\
9 & \pro & \D(\D(\D(\c{B_{4}}, \D(7, \D(\D(\D(\c{B_{4}}, \D(\D(4, 8), 7)), 3), 8))), 6), 8)\\
10 & \pro & \D(\D(\D(\c{B_{4}}, 9), 9), 9)\\
\end{array}$

\subsubsection{(3) Final Converted Proof from the Original Axioms}\ \smallskip

\begin{tabular}{l@{\hspace{1em}}l}
  Compacted size & 50\\
  Tree size & 1,541\\
  Height & 40\\
\end{tabular}

\medskip

$\begin{array}{rcl}
4 & \pro & \D(1, 1)\\
5 & \pro & \D(\D(2, \D(1, 2)), 1)\\
6 & \pro & \D(\D(2, \D(1, \D(\D(2, 2), 4))), \D(\D(2, 4), 2))\\
7 & \pro & \D(\D(2, \D(1, \D(2, \D(1, 5)))), 5)\\
8 & \pro & \D(7, 3)\\
9 & \pro & \D(\D(8, 3), 1)\\
10 & \pro & \D(\D(6, \D(\D(\D(7, 1), 3), \D(2, \D(\D(8, \D(2, 9)), 1)))), 1)\\
11 & \pro & \D(2, \D(1, 10))\\
12 & \pro & \D(\D(2, \D(1, \D(\D(6, \D(\D(\D(7, 11), 2), 1)), 9))), 2)\\
13 & \pro & \D(\D(\D(7, \D(11, \D(\D(\D(7, \D(\D(8, 12), 11)), 3), 12))), 10), 12)\\
14 & \pro & \D(\D(\D(7, 13), 13), 13)\\
\end{array}$

\end{showproof}

\clearpage
\section{Table with Result Data for All \POI Theorems}
\label{app-bigtable}

The following table shows for each \POI theorem the times in seconds that were
required for solving the corresponding proving problem, as described in
Sect.~\ref{sec-ideal}. Column \h{Rtg} shows the theorem's rating as defined in
Sect.~\ref{sec-ideal}. Column \h{Best} shows the minimum of the time values
for the first-order provers. The token ``--'' indicates timeout.
Column~\h{Refs} shows the number of references to the theorem in the proofs of
other theorems in the considered first two thirds of \SETMM. The rows are
ordered lexically by the following criteria: \h{Rating} (high-to-low),
\h{Best} (high-to-low), \h{Refs} (low-to-high), theorem name (alphabet).
Column \h{Status} indicates whether a proof of the theorem was obtained using
the presented techniques with the following keys:
\begin{itemize}
\item \STP: A proof of the theorem was obtained in some base run of \SGCD with
  PSP level characterization ($\RBP$, $\RBPF_{8-12}$, see
  Sect.~\ref{sec-basesolved}).
\item \STT: A proof of the theorem was obtained in the base run of \SGCD with
  tree size level characterization ($\RBT$, see Sect.~\ref{sec-basesolved}).
\item \STS: A proof of the theorem was obtained by a run of \SGCD with lemma
  synthesis via proof term compression ($\RS_{1-13}$, see
  Sect.~\ref{sec-dag-synthesis}).
\item \STC: A proof of the theorem was obtained by a run of \SGCD involving
  combinatory proof schemas ($\RC_{1-22}$, see Sect.~\ref{sec-combinatory}).
\item \STL: The proving problem was solved in 3,600~s with \Vampire, if the
  input is enhanced with our lemmas from experiment $\RS_3$
  (see Sect.~\ref{sec-lemmas-fo-provers}).
\end{itemize}

{\scriptsize \centering
\begin{longtable}{ll@{\hspace{8pt}}rrrrrrrr}
  \mm{Theorem} & \h{Status} & \h{Rtg} & \h{Best} &  \Vampire & \EProver &  \EProverAS &  \ProverN & \leanCoP &  \h{Refs}\\\midrule
  \endhead
 \mm{stoic4b}         & \STE\STE\STE\STE\STL & 1.00 &      -- &      -- &      -- &      -- &      -- &      -- &         0\\
\mm{cadifp}          & \STE\STE\STE\STE\STL & 1.00 &      -- &      -- &      -- &      -- &      -- &      -- &         0\\
\mm{cadnot}          & \STE\STE\STE\STE\STE & 1.00 &      -- &      -- &      -- &      -- &      -- &      -- &         0\\
\mm{cadrot}          & \STE\STE\STE\STE\STE & 1.00 &      -- &      -- &      -- &      -- &      -- &      -- &         0\\
\mm{cadbi123i}       & \STE\STE\STE\STE\STE & 1.00 &      -- &      -- &      -- &      -- &      -- &      -- &         0\\
\mm{hadifp}          & \STE\STE\STE\STE\STE & 1.00 &      -- &      -- &      -- &      -- &      -- &      -- &         0\\
\mm{hadbi123i}       & \STE\STE\STE\STE\STE & 1.00 &      -- &      -- &      -- &      -- &      -- &      -- &         0\\
\mm{norass}          & \STE\STE\STE\STE\STL & 1.00 &      -- &      -- &      -- &      -- &      -- &      -- &         0\\
\end{longtable}
}

\end{document}